\documentclass{JHEP3}

\input{epsf}
\usepackage{epsfig}
\usepackage{amssymb}
\usepackage{amsfonts}
\usepackage{amsbsy}

\def\IIb{{IIb}}

\def\adscc{{\hat\Lambda}}

\def\im{\mbox{Im }}

\def\IC{\mathbb{C}}

\def\IZ{\mathbb{Z}}
\def\IR{\mathbb{R}}

\def\CM {{\cal M}}
\def\CN {{\cal N}}

\def\CF {{\cal F}}

\def\CZ {{\cal Z}}

\def\CC {{\cal C}}
\def\CB {{\cal B}}

\def\CK{{\cal K}}
\def\CZ{{\cal Z}}
\def\la{\langle}
\def\ra{\rangle}

\def\tM{\tilde{M}}

\newcommand{\eq}[1]{Eq.~(\ref{eq:#1})}

\renewcommand{\Im}{{\rm Im }}
\renewcommand{\Re}{{\rm Re }}

\def\one{{\hbox{ 1\kern-.8mm l}}}

\def\tr{{\rm tr\,}}
\def\sgn{{\rm sgn\,}}

\def\p{\partial}
\def\ba{\bar{a}}
\def\bb{\bar{b}}
\def\bc{\bar{c}}
\def\bd{\bar{d}}
\def\be{\bar{e}}
\def\bz{\bar{z}}
\def\bZ{\bar{Z}}
\def\bW{\bar{W}}
\def\bD{\bar{D}}

\def\bB{\bar{B}}

\def\bPi{\bar{\Pi}}
\def\bOmega{\bar{\Omega}}
\def\bpartial{\bar{\partial}}
\def\bj{{\bar{j}}}
\def\bi{{\bar{i}}}
\def\bk{{\bar{k}}}
\def\bl{{\bar{l}}}
\def\bm{{\bar{m}}}

\def\bpsi{\bar{\psi}}

\def\bt{\bar{t}}
\def\bv{\bar{v}}

\def\bz{\bar{z}}
\def\btau{\bar{\tau}}

\def\bB{\bar{B}}

\def\bX{\bar{X}}
\def\bY{\bar{Y}}
\def\bZ{\bar{Z}}

\def\un{\underline{0}}

\title{Distributions of flux vacua}

\author{Frederik Denef$^{1,2}$ and Michael R. Douglas$^{1,\&,2}$\\
$^1$NHETC and Department of Physics and Astronomy\\
Rutgers University\\
Piscataway, NJ 08855--0849, USA\\
\\
$^\&$I.H.E.S., Le Bois-Marie, Bures-sur-Yvette, 91440 France\\
\\
$^2$Caltech, Pasadena CA 91125, USA\\
{\tt denef, mrd@physics.rutgers.edu}
}

\abstract{
We give results for the distribution and number of flux vacua of various
types, supersymmetric and nonsupersymmetric, in \IIb\ string theory
compactified on Calabi-Yau manifolds.  We compare this with related problems
such as counting attractor points.
}

\begin{document}

\section{Introduction}

In this work, we study the distribution of metastable supersymmetric
and nonsupersymmetric flux vacua in Calabi-Yau compactification of
various string theories, along the lines developed in the works
\cite{stat,AD,DSZ}

The effect of turning on gauge field strengths (or ``flux'') in string
compactification has been studied in many works, starting with
\cite{Strom}.  Some recent examples include
\cite{Giryavets:2003vd,Grimm:2004uq}.  Perhaps the most important
qualitative effect of flux is that, since its contribution to the
energy depends on the moduli of the compactification manifold,
minimizing this energy will stabilize moduli, eliminating undesired
massless fields.  Since coupling constants in the low energy theory
depend on moduli, finding the values at which moduli can be stabilized
is an essential step in determining low energy predictions.  It has
also been suggested that taking into account the large number of
possible choices for the flux, will lead to a large number of vacua
with closely spaced values of the cosmological constant, and that some
of these will reproduce its small observed value just on statistical
grounds \cite{BouPol}.  Thus one would like to know the distribution
of cosmological constants, and how this depends on the moduli and
other parameters of the vacuum.

There are a lot of flux vacua, and finding each one explicitly is a
lot of work.  Furthermore, we are not entirely sure what properties we
seek: there are many different scenarios for string phenomenology,
each requiring different properties of the vacuum.  Thus, rather than
study individual vacua, we believe it is more interesting at this
point to study the overall distribution of vacua in moduli space, and
the distribution of quantities such as the cosmological constant and
supersymmetry breaking scale.  As discussed in \cite{stat,durham}, such
statistical results can serve as a guide to string phenomenology, and
provide a ``stringy'' definition of naturalness.  And, they are not much
harder to get than results for individual vacua, as was seen in
\cite{AD,DSZ} and as we will demonstrate here.

A useful way to state these problems is as that of finding vacua in a
specific ensemble (or set) of $\CN=1$ effective supergravity theories,
for which the K\"ahler potential and superpotential can be found
explicitly.  In principle, these theories are obtained by listing all
string/M theory compactifications in a certain class, and in each case
integrating out all but a finite number of fields, to obtain a
Lagrangian valid at a low energy scale $E$.  While not all vacua can
be described by effective field theory, since at energies studied so
far our universe seems to be described by effective field theory, this
restriction seems adequate for the basic physics we want.

One can certainly question whether this type of analysis captures all
consistency conditions which vacua must satisfy.  Perhaps the most
important examples would be stability over cosmological time scales,
and higher dimensional consistency conditions which are not obvious
after integrating out fields.  It is entirely possible that there are
others; a list of speculations in this direction appears in
\cite{Banks}, and these deserve study.  Furthermore, it might turn out
that early cosmology selects or favors a subset of preferred vacua.
Our philosophy is {\bf not} that we believe that none of this is
important and thus can put absolute trust in the vacuum counting
results below.  Rather, we believe that, even if we had this
additional information, it would not tell us which vacuum to consider
{\it a priori}, and we would still need to make an analysis of the
type made here to find the relevant vacua, with the additional
information taken into account as well.  Thus the results we give
should be considered as formal developments, with suggestive
implications for real physical models, but which might be modified in
light of better understanding.

While the techniques we will describe could be used for any explicit
ensemble of effective supergravity theories, we work here with
Calabi-Yau compactification in the large volume, weak coupling limit,
because good techniques for explicitly computing and working with the
resulting supergravity Lagrangians exist at present only for this
case.  Indeed, type \IIb\ vacua on Calabi-Yau might turn out to be
``representative'' in the sense discussed in \cite{stat,durham}, on
various grounds.  First, known dualities relate many other
nonperturbative superpotentials to these cases, and it seems fair to
say that all the structure in the potential which has been called on
in model building so far, such as generation of exponentially small
scales, and spontaneous supersymmetry breaking, can be seen in flux
superpotentials.  Second, dualities have been proposed which relate
many of the other large classes of vacua to these.  A systematic way
to study the hypothesis that (say) \IIb\ on Calabi-Yau is
representative, would be to find statistics of vacua from {\bf two} or
more large sets of constructions; if both were representative, clearly
these statistics would have to be the same.  The present results are a
necessary first step towards such a test, namely to find statistics
for {\bf one} large set of constructions.

This concludes the justification of our approach.  Our main discussion
is somewhat technical, so we devote the remainder of the introduction
to a basic overview of the type of results we will get.

\subsection{Distributions of vacua}

Our starting point is to imagine that we are given a list of
effective supergravity theories $T_1$, $T_2$, etc. all with the
same configuration space (the space in which the chiral fields
take values).  We consider here theories with no gauge sector, so
a theory $T_i$ is specified by a Kahler potential ${\cal K}_i$ and
superpotential $W_i$.

We then apply the standard $\CN=1$ supergravity formula for the
potential,
\begin{equation} \label{eq:defV}
V = e^{{\cal K}/M_p^2}\left( g^{i\bj} D_iW D_{\bj} W^* -
 {3\over M_p^2} |W|^2 \right) + D^2,
\end{equation}
and look for solutions of $\p V/\p z^i = \p V/\p \bz^\bi=0$.
Here $M_p$ is the four dimensional Planck scale (which will shortly
be set to $1$).

Vacua come in various types.
First, as is familiar, a supersymmetric vacuum is a solution of
\begin{equation} \label{eq:susyvac}
D_iW(z) = \p_i W + {1\over M_p^2}(\p_i {\cal K}) W = 0 .
\end{equation}
We consider both Minkowski $W=0$ and AdS $W\ne 0$ vacua.

Conceptually, the simplest distribution we could consider is
the ``density of supersymmetric vacua,'' defined as
$$
d\mu_s(z) = \sum_{i} \delta_z(DW_i(z))
$$
where $\delta_z(f)$ is a delta function at $f=0$, with a
normalization factor such that each solution of $f=0$ contributes unit
weight in an integral $\int d^{2n} z$.  Thus, integrating this density
over a region in configuration space, gives the number of vacua which
stabilize the moduli in this region.  A mathematically precise
definition, and many explicit formulae which can be adapted to the
physical situation, can be found in \cite{DSZ}.

In this case,
$$
\delta_z(DW(z)) \equiv \delta^{(n)}(DW(z)) \delta^{(n)}(\bar DW^*(\bz))
 |\det D^2W(z)|
$$
where the Jacobian term is introduced to cancel the one arising from
the change of variables $z\rightarrow DW$.  The matrix $D^2W$ is
a $2n\times 2n$ matrix
\begin{equation}\label{eq:Jacobian}
D^2W \equiv \left(\matrix{\bar\partial_\bi D_j W(z)& \partial_i
D_j W(z)\cr \bar\partial_\bi \bar D_\bj W^*(z)& \partial_i \bar
D_\bj W^*(z)}\right).
\end{equation}
Essentially, this is the fermionic mass matrix. Note that we could
have replaced the partial derivatives by covariant derivatives, as
$D_i D_j W = \partial_i D_j W$ when $DW=0$.

With this definition, a vacuum with massless fermions counts as zero.
There are better definitions, discussed in \cite{stat}, which would be
appropriate if generic vacua had massless fermions.  However, since
generic vacua in our problem are isolated and have no massless
fermions, this definition is fine.

We can also define joint distributions such as the distribution of
supersymmetric vacua with a given cosmological constant, $$
d\mu_s(z,\Lambda) = \sum_{i} \delta_z(DW_i(z)) ~
\delta(\Lambda-(-3e^{{\cal K}_i}|W_i(z)|^2)) . $$ Below, we will
define similar densities for nonsupersymmetric vacua of various
types; at this point the basic idea should be clear.

\subsection{Approximate distributions of vacua}

Now, if we have a finite list of supergravity theories $T_i$, and if
in each the number of vacua is finite, such a density will be a sum
of delta functions.  This is hard (though not impossible) to study, and
for many purposes one might be satisfied with a continuous approximation
to this density, a function $\rho(z)$ whose integral over a region $R$,
\begin{equation} \label{eq:region}
\int_R d^{2n}z \rho(z) ,
\end{equation}
approximates the actual number of vacua in this region.

What does this mean and what good is it?  To give the question some
context, suppose we had an explicit string theory construction of the
Standard Model, and we were trying to decide whether it could
reproduce the gauge and Yukawa couplings.  While in some cases these
are constrained by symmetry, this is not enough to determine the
non-zero couplings.  In the vast majority of explicit models, these
couplings depend on moduli of the compactification (metric, bundle
and brane moduli, etc.)  and most results in this area address the
problem of finding the formula for the couplings in terms of moduli.

Let us suppose we have such a formula.  We could then use it to
identify a region $R$ in moduli space, or a region in the joint space
of moduli, cosmological constant and other observables, such that any
vacuum in this region would be guaranteed to produce couplings which
agreed with observation to the required precision (let us say such
vacua ``work'').  Thus the question would become, does the region $R$
contain vacua, which can be obtained by stabilizing moduli.

Now, suppose we had an approximation $\rho(z,\Lambda)$ to the density
of suitable vacua, with suitably small cosmological constant.
Integrating it over the region $R$, would produce an approximation to
the number of vacua which work.  This starts to sound interesting, but
of course we do not really want an approximate answer.  In the end, we
want to know if the approximation helps to answer the real question of
finding {\bf a} vacuum which works.

If the region is small, or the density of vacua is small, one might
need to interpret a result such as ``approximately $10^{-20}$
vacua work.''  For the results we will discuss, this basically means
that one expects that no vacua work, but it is possible that
structure not reproduced by the approximation or some chance fine tuning
will nevertheless lead to the existence of vacua.
If there are competing classes of vacua which work, this would start
to be evidence that vacua in the class under study do not work.
To develop this hypothesis, the next question
would be, what more do we need to do to {\bf prove} that this region
contains no vacua.

If the region $R$ is large enough to contain many vacua, there are two
cases worth distinguishing.  In our present state of ignorance, what
we can typically do is try to enforce some but not all of the
observational constraints on our model, and thus the number we would
get at this stage would just be one factor in a final result.  In this
case, the most natural condition to put is that the integral
\eq{region} should approximate the actual number of vacua $N(R)$ with
an error much less than $N(R)$.

Suppose we have solved the problem to the end, or we feel that the
vacuum we have found is particularly interesting, say because it
realizes some property of interest or refutes some conjecture in the
literature.  Then we would like to use our calculation of $N(R) >> 1$
to {\bf prove} that a vacuum for the original, unapproximated problem,
indeed sits in $R$; for a good approximation, this will be possible.

Of course, we might decide that $N(R)$ is so large, that the original
goal of the discussion, say to test whether this class of vacua
can reproduce the couplings in the Standard Model,
becomes pointless.  One can of course still hope that the assumptions that
went into our choice of supergravity theories and definition of vacuum
are false; however the computation of $N(R)$ under these assumptions would
have been good enough and one would not need to improve that.

This covers the various possibilities.  The main points we want to
make here are the following.  First, what one wants to know next, and
how one thinks about the problem, depends very much on whether $N(R)
>> 1$, $N(R) \sim 1$ or $N(R) << 1$, which is thus as important a
question as finding particular vacua.

Second, given that we trust our definition of ``vacuum,'' it is
actually less important to know how a given vacuum is obtained ({\it
e.g.}  by which choice of flux) than to know that it exists and
stabilizes moduli (if the observables are controlled by the moduli,
not by the fluxes; of course the fluxes could appear explicitly in the
observables as well).

Having said all this, we defer the discussion of how one could
proceed in these various cases, to section 5 and
to other work to appear.

\subsection{Other types of vacuum}

There are two types of nonsupersymmetric vacua we consider.  The
simpler possibility is breaking due to non-flux effects.  In other
words, we still seek solutions of \eq{susyvac}, calling upon other
effects to break supersymmetry and lift the potential energy.  This
was invoked, for example, in \cite{KKLT}, which proposed to break
supersymmetry by adding an anti D3-brane in \IIb\ compactification.
Another possibility is to call on D term supersymmetry breaking, as
has been discussed in many works.  Indeed, to the extent the breaking
can be understood in terms of effective $\CN=1$ supergravity, this is
the only possibility, and there are arguments in the literature that
supersymmetry breaking by adding antibranes or misaligned branes is of
this type \cite{Dbranes,Aspinwall:2001dz,Binetruy:2004hh}.
Thus, we are going to refer to this as D-type breaking.

Before considering stability, the distribution of D-type vacua is the
same as that for supersymmetric vacua, up to a factor which expresses
the fraction of vacua which allow non-flux supersymmetry breaking.
This might depend on the example at hand; we will simply set it to
$1$ here.

Granting that in a given vacuum, adding the
supersymmetry breaking term results in zero cosmological constant,
we would identify the supersymmetry breaking scale as
$$
M_{susy}^4 = 3 \adscc
$$
where
\begin{equation}\label{eq:AdScc}
\adscc = e^{{\cal K}(z,\bar z)}|W(z)|^2
\end{equation}
is the norm of the superpotential for a vacuum stabilized at $z$.
Of course, without the supersymmetry breaking, the cosmological
constant of the resulting vacuum would have been $\Lambda_{AdS}=-3\adscc$,
so we will often refer to $\adscc$
as the ``AdS cosmological constant.''

The other type of nonsupersymmetric vacuum is pure F type
breaking; in other words to find a solution of $V'=0$ which is not
a solution of \eq{susyvac}.  The scale of this breaking is given
by $$ M_{susy}^4 = e^{\cal K} g^{i\bj} D_iW D_\bj W^* $$ which for
$V=0$ is equal to the above.  The density of these vacua is given
by $$ d\mu_F(z) = \sum_i \delta_z(V'(z)) . $$

In general, one can have mixed D and F breaking.  This is interesting
only when both D and F terms depend on the same fields, which can
only come about from non-flux effects, and is thus beyond our scope here.

In any case, the most interesting nonsupersymmetric vacua are
the metastable (tachyon free) vacua, with $V''$ positive definite.
This constraint can be enforced formally by definitions such
as
 $$
 d\mu_{F,metastable}(z) = \sum_i \delta_z(V'(z)) \theta(V''(z))
 $$
where $\theta(V'')$ is $1$ when the $2n\times 2n$ real matrix of
squared bosonic masses $M=V''$ is positive definite.
The derivatives of $V$ appearing here are
\begin{eqnarray} \label{eq:Vderivs}
 \partial_a V &=& e^{\cal K} ( D_a D_b W \bD^b \bW - 2 D_a W \bW ) \label{eq:DV} \\
 D_a \partial_b V &=& e^{\cal K} (D_a D_b D_c W \bD^c \bW - D_a D_b W \bW) \label{eq:DDV}\\
 \bD_{\ba} \partial_b V &=&
 e^{\cal K}({R^d}_{c \ba b} D_d W \bD^c \bW
 + g_{b\ba} D_c W \bD^c \bW - D_b W D_{\ba} \bW \nonumber \\
  && - 2 g_{b\ba} W \bW
 + D_b D_c W \bD_{\ba} \bD^c \bW), \label{eq:DcDV}
\end{eqnarray}
where $R$ is the curvature of the cotangent bundle, i.e.
${R^d}_{ca \bb} \, X_d \equiv [\nabla_a,\bar{\nabla}_{\bb}] X_c =
\bpartial_{\bb}(g^{\be d}
\partial_a g_{c \be}) \, X_d$. Note that we could have replaced
the covariant derivatives by ordinary partial derivatives, because
$D d V = d^2 V$ when $dV=0$.

For D breaking, positivity $M>0$ can be
analyzed as follows. First observe that in general, if $DW=0$,
\begin{equation} \label{eq:MHH}
 M = H^2 - 3 \adscc^{1/2} H,
\end{equation}
where
\begin{equation}
 H = 2 \, d^2 \adscc^{1/2}.
\end{equation}
This follows directly from \eq{DDV} and \eq{DcDV}. Thus, to have
$M>0$, all eigenvalues $\lambda$ of $H$ must satisfy $\lambda < 0$
or $\lambda > 3 \adscc^{1/2}$. In particular, if $W=0$ at the critical
point, $M$ is automatically non-negative, and by continuity the
same will be true for most susy vacua with small $\adscc$.
On the other hand,
small positive eigenvalues of $H$ will lead to tachyons and instability.

The actual computations of all of these densities will of course
rely heavily on specific details, but a general point worth
keeping in mind is that {\bf any} joint density $$ d\mu(z,a_n) =
\sum_i \delta_z(V'(z))~ \delta(a_n - O_n(z)) $$ of vacua in moduli
space along with any observables $O_n(z)$ defined in terms of the
Taylor series expansion of the effective Lagrangian about the
vacuum (masses, couplings of moduli, etc.) can be computed if we
simply know the joint distribution of $W(z)$, ${\cal
K}(z,\bar{z})$ and a finite number of their derivatives evaluated
{\bf at} the point $z$, in other words a finite number of
variables. Although obvious, this is very useful in structuring
the problem, and is the main reason why this class of problem is
so much simpler than problems involving flows on the moduli space.

\subsection{Index densities}

Finally, there is a quantity we call the ``index density''.  In the
particular case of supersymmetric vacua, it is
$$
d{\cal I}_s(z) = \sum_{i} (-1)^F \delta_z(DW_i(z))
$$
where the index $(-1)^F$ of a vacuum stabilized at $z$ is defined to be
\begin{equation}\label{eq:index}
(-1)^F \equiv \sgn \det_{i,j} D^2W(z)
\end{equation}
with $D^2W(z)$ as defined in \eq{Jacobian}.

The simplest reason to consider this is that the index is precisely the
sign of the Jacobian which appeared in defining $\delta_z(DW)$, so
$$
(-1)^F \delta_z(DW(z)) \equiv \delta^{(n)}(DW(z)) \delta^{(n)}(\bar DW^*(\bz))
 \det D^2W(z)
$$
with no absolute value signs.  Thus it is easier to compute, and
provides a lower bound for the actual number of vacua.  A formula for
the index, and the explicit result for $T^6/\IZ_2$ compactification,
were given in \cite{AD}.

There are also conceptual reasons to be interested in the supergravity
index, as discussed in \cite{Strings2002,stat}.  To start,
let us first comment on some differences between the problem of
finding vacua in supergravity, and
the much better studied problem of finding vacua in globally
supersymmetric theories, satisfying
\begin{equation} \label{eq:globalvac}
\p_i W(z) = 0 .
\end{equation}
Of course,  \eq{susyvac} reduces to this upon taking the limit
$M_{pl}\rightarrow \infty$, or equivalently if all structure in
$W$ is on scales much less than $M_{pl}$ (assuming derivatives of
${\cal K}$ do not grow with $M_{pl}$).  On the other hand, we need
the supergravity correction to interpolate between different field
theoretic limits.  Of course, all hopes for getting a small
cosmological constant out of \eq{defV} rest on the supergravity
term $-3|W|^2$ as well.

The problem of finding solutions of \eq{globalvac} is holomorphic and
therefore much easier than for \eq{susyvac}.  In particular, vacua
cannot be created or destroyed under variation of parameters, they
can only move off to infinity or merge together.  This makes it
possible to give topological formulae for the total number of vacua
in global supersymmetry; the possibility of vacua merging is accounted
for by counting such vacua with multiplicity.

Can we do the same for supergravity vacua? Evidently not, because
one can construct a family of Kahler potentials ${\cal K}_t(z)$
such that varying $t$ creates pairs of solutions of \eq{susyvac}.
One way to see this is to note that, in a region in which $W\ne
0$, \eq{susyvac} is equivalent to the condition that we are at a
critical point of the function $\adscc$ from \eq{AdScc}. Thus,
where $W \ne 0$, one can apply Morse theory to this problem, as
discussed in \cite{Behrndt:2001qa}. It is well known in this
context that critical points can be created and destroyed in
pairs.

Clearly we cannot hope for a topological formula for the total
number of vacua, but the above suggests using the Morse index for
$\Lambda_{AdS}$ as a lower bound for the number of vacua, which
might admit a topological formula.  However, this is not correct
because critical points of $\Lambda_{AdS}$ are not necessarily
vacua; indeed every point with $W=0$ (and ${\cal K}$ nonsingular)
is a critical point of $\Lambda_{AdS}$.

The search for a topological formula runs into other difficulties as
well.  Most importantly, the configuration spaces which appear in
known examples of effective supergravities are not compact, and cannot
be compactified.  The prototypical example is the upper half plane.
This configuration space has a boundary, the real axis, and it is
easy to see in examples that varying parameters (e.g. flux)
can move vacua in and out of the configuration space.  Furthermore,
different flux sectors can contain different numbers of
vacua.\footnote{This statement is a bit imprecise; a more precise
explanation taking duality into account is given in \cite{AD}.}

Anyways, the correct generalization of the Morse index to this
situation, as discussed in \cite{Strings2002,DSZ}, is \eq{index}.
This agrees with the Morse index when $W\ne 0$, and is $(-1)^n$ when
$W=0$.\footnote{In \cite{Strings2002,AD} we instead used a
convention for the index which includes an extra factor of $(-1)^n$,
so that Minkowski vacua always count $+1$.  There are arguments in favor
of both conventions, and one should be careful to note which is in use.}

Since $\adscc=e^{\cal K}|W|^2$ is a general (non-holomorphic)
function, which away from $W=0$ can be deformed fairly arbitrarily
by deforming ${\cal K}$, there is no obvious reason that one could
not deform it to remove all cancelling pairs of
vacua.\footnote{The function ${\cal K}$ must satisfy the
constraint that $\partial\bar\partial {\cal K}$ is positive
definite, but in one dimension this does not seem to prevent
deforming away pairs of vacua.}  Thus, the index is the absolute
minimal number of vacua which could be obtained by deforming the
K\"ahler potential.

Thus, one physical way to think about the index, is to say that the
difference between the actual number of vacua, and the index, in some
sense measures the number of ``K\"ahler stabilized vacua,'' vacua
whose existence depends on both the superpotential and the K\"ahler
potential.  We found in \cite{DSZ} and will find below that the actual
density of vacua is typically the index density times a bounded
function greater than one, so in this precise sense, there are many
K\"ahler stabilized vacua.

Since the $D\bar D$ terms in \eq{Jacobian} go away in the limit
$M_{pl}\rightarrow \infty$, the index of a vacuum which survives this limit,
and thus is not ``K\"ahler stabilized,''
will necessarily be $(-1)^n$ (the same as for $W=0$).  Conversely, one
could say that the vacua with index of the opposite sign are all
K\"ahler stabilized, and would go away in this limit.  One should
realize however that this limit is highly ambiguous (the results
change under K\"ahler-Weyl transformation) and that it may not in
general make sense to say which particular vacua with index $(-1)^n$
are K\"ahler stabilized or not.

\subsection{Ensembles of flux vacua}

We next discuss the set or ensemble of vacua we consider.
General arguments have been given to the effect that the large
volume, weak coupling limit of compactification
of string/M theory on a Ricci-flat manifold $M$ with flux, can be described
by a $d=4$ effective supergravity Lagrangian, whose configuration space
$\CC$ is the moduli space of compactifications on $M$ with no flux,
the K\"ahler potential is taken to be the one for zero flux, and
whose superpotential is the Gukov-Vafa-Witten superpotential
\cite{Gukov:1999ya}, which
takes the form
\begin{equation}\label{eq:gvw}
W = \int_M  G \wedge \Omega(z) ,
\end{equation}
where $z$ are the complex structure moduli, $\Omega(z)$ is an
appropriate form (depending on the theory and $M$), and $G$ is the
$p$-form gauge field strength, which we normalize to have integral
periods.

Although we will discuss flux compactification of various
theories: F theory on fourfolds, heterotic string on CY$_3$, and
the (formally very similar) attractor description of black hole
entropies, we work mostly with the \IIb\ flux compactifications on
Calabi-Yau developed by Giddings, Kachru and Polchinski
\cite{Giddings:2001yu}. Then $\Omega$ is the holomorphic
three-form on the CY, and $G$ is a sum of the NS and RR three-form
gauge field strengths $$ G = F^{RR} - \tau H^{NS} , $$ with
$F,H\in H^3(M,\IZ)$, and $\tau=C^{(0)}+ie^{-\phi}$ is the
dilaton-axion.

The ``ensemble of effective field theories'' with flux is then the
set of supergravity theories with $W$ given by \eq{gvw}, with $G$
satisfying the tadpole constraint
 $$ \int F^{RR} \wedge H^{NS} \leq L_*.$$
This is discussed in much more detail in \cite{AD}.

We will leave out the K\"ahler moduli and forget about their contribution
to the potential, again for the reasons discussed in \cite{AD}.  We
are going to derive many results relevant for this part of the problem,
and will discuss it a bit in the conclusions, but reserve most of what
we have to say about it to other work. \cite{DDF}.

With the above definition of $W$, the F-term potential $V$ is
given by \cite{Giddings:2001yu}:
\begin{equation}
 V = 2 T_3 e^{\cal K} (D_a W \bD^{a} \bW - 3 W \bW),
\end{equation}
where $T_3=(2\pi)^{-3} {\alpha'}^{-2}$ is the D3-brane tension (in
physical units). In subsequent sections we set $2 T_3 \rightarrow 1$, not
as a choice of units (since $M_{pl}=1$), but rather by
shifting ${\cal K}$ by a constant.

Since $M_{Pl,4}^2 = V_6 M_{Pl,10}^8$, the dimensionless ratio
$V/M_{Pl,4}^4 \sim 1/V_6^2$, but this is only because $M_{Pl,4}$
grows with $V_6$.  For orientation, in the traditional KK
scenarios, $V_6 \sim l_s \sim 1/M_{Pl,10}$ up to $O(1)$ factors,
so the natural energy scale of the flux potential is the string
scale, and we will want supersymmetry breaking at scales $M_{susy}
l_s<1$ or even $<<1$. In a ``large extra dimensions'' scenario,
$V_6/l_s^6 >> 1$, and we might accept $M_{susy} l_s \sim 1$.

Our basic results are obtained by neglecting the quantization of flux.
This is expected to be a good approximation in the limit that the flux
is large compared to other numbers such as the number of cycles.  We
will discuss this limit, and the sense in which the smooth
distribution approximates the distributions of vacua at finite $L$, in
section 5.  Our tools for doing this will be number theoretic theorems
which state conditions on a ``region in flux space'' which guarantee
that its volume provides a good estimate for the number of lattice
points it contains.

\section{Notations and some useful formulas}

To avoid dragging along factors of $e^{\cal K}$ in the
calculations, we will slightly change notation in what follows and
denote the usual holomorphic superpotential by $\hat{W}$ and
reserve $W$ for the K\"ahler invariant normalized but
non-holomorphic superpotential:
\begin{equation}
 W(z,\bar{z}) = e^{{\cal K}(z,\bar{z})/2} W(z).
\end{equation}
Similarly we write
\begin{equation}
 \Omega(z,\bar{z}) = e^{{\cal K}(z,\bar{z})/2} \hat{\Omega}(z)
\end{equation}
for the normalized holomorphic form on the Calabi-Yau. We modify
the definition of the covariant derivative accordingly: $D_a W
\equiv e^{\CK/2} D_a \hat W$, etc.

Consider first a general F-theory flux compactification on an
elliptically fibered Calabi-Yau fourfold $X$. The
Gukov-Vafa-Witten flux superpotential is
\begin{equation} \label{eq:fluxW}
 W=\int_X G_4 \wedge \Omega = N^\alpha \Pi_\alpha,
\end{equation}
where the $\Pi_\alpha = \int \Sigma_\alpha \wedge \Omega$ are the
periods of some basis $\{ \Sigma_\alpha \}$ of $H^4(X,\IZ)$. We
normalize $G_4$ such that $G_4 \in H^4(X,\IZ)$,\footnote{The flux
quantization condition can actually be shifted by a nonintegral
constant in some circumstances \cite{Witten:1996md}, but since we
will make a continuum approximation for the fluxes anyway, we can
ignore such subtleties. Also, in F-theory, not all fluxes in
$H^4(X,\IZ)$ are allowed: essentially, one leg should be on the
elliptic fiber \cite{Gukov:1999ya}.} so $N \in \IZ^{b_4}$. The
flux has to satisfy the tadpole cancellation condition
\begin{equation}
 L \equiv \frac{1}{2} \int_X G_4 \wedge G_4 = \frac{\chi(X)}{24} -
 N_{D3}
\end{equation}
where $N_{D3}$ is the number of D3 branes minus the number of
anti-D3 branes transversal to $X$. If we are looking for
supersymmetric vacua, this gives an upper bound
\begin{equation}
 L \leq L_*
\end{equation}
with $L_* = \chi(X)/24$. Allowing anti-D3 branes, $L_*$ can become
bigger, but not indefinitely, as a sufficient number of anti-D3
branes in a flux background will decay into a state with flux and
D3 branes only \cite{Kachru:2002gs,DeWolfe:2004qx}.

The superpotential depends on the complex structure moduli $z^a$
($a=1,\ldots,h^{3,1}(X)$) only. The metric on complex structure
moduli space is the Weil-Petersson metric, derived from the
K\"ahler potential
\begin{equation}
 \CK = -\ln \la \hat{\Omega} , \hat{\bOmega} \ra
 \equiv - \ln \int_X \hat{\Omega} \wedge \hat{\bOmega}
 = - \ln \hat{\Pi}_{\alpha} (\eta^{-1})^{\alpha\beta}
 \hat{\bPi}_{\beta},
\end{equation}
where $\eta_{\alpha\beta}$ is the intersection form with respect
to the basis $\{\Sigma_\alpha\}$.

We will often encounter intersection products of the covariant
derivatives of the normalized period vector (or holomorphic
4-form):
\begin{eqnarray}
 \CF_{A \ldots B | C \ldots D}
 &\equiv& D_A \cdots D_B \Pi \, \eta^{-1} \, D_C \cdots D_D \bPi
 = \la D_A \cdots D_B \Omega , D_C \cdots D_D \bOmega \ra \\
 \CF'_{A \ldots B | C \ldots D}
 &\equiv& D_A \cdots D_B \Pi \, \eta^{-1} \, D_C \cdots D_D \Pi
 = \la D_A \cdots D_B \Omega , D_C \cdots D_D \Omega \ra,
\end{eqnarray}
where the capital indices can be either holomorphic or
anti-holomorphic. These are most easily calculated by using
identities of the form $\la DX, Y \ra = D\la X,Y \ra - \la X, DY
\ra$, orthogonality of $(4,4-k)$ and $(4-k',4)$-forms with $k'\neq
k$, and commutation relations of $D$ and $\bD$, together with
Griffiths transversality, i.e.\ acting with $k$ derivatives on the
$(4,0)$-form $\Omega$ gives a sum of $(4-q,q)$-forms with $q$ at
most equal to $k$. In fact, $D_a\Omega$ is pure $(3,1)$ and $D_a
D_b\Omega$ is pure $(2,2)$, which can be shown in similar fashion.
As an example, we have $\CF_{a|\bb}=\la D_a \Omega,\bD_{\bb}
\bOmega \ra = D_a \la \Omega,\bD_{\bb} \bOmega \ra - \la
\Omega,D_a \bD_{\bb} \bOmega \ra = 0 - \la \Omega,g_{a\bb} \bOmega
\ra = -g_{a\bb}$. Similarly, the absence of a (3,1)-part in $D_a
D_b \Omega$ is follows from $\CF_{ab|\bc} = \la D_a D_b
\Omega,\bD_{\bc} \bOmega \ra = D_a \la D_b \Omega,\bD_{\bc}
\bOmega \ra - \la D_b \Omega, D_a \bD_{\bc} \bOmega \ra = - D_a
g_{b\bc} - \la D_b \Omega, g_{a\bc} \bOmega \ra = 0$. Thus, most
lower order $\CF$-tensors vanish. Some nonzero ones are
\begin{eqnarray}
 \CF_{a|\bb} &=& -g_{a \bb} \\
 \CF_{ab|\bc\bd} &=& R_{a\bc b\bd} + g_{b\bd} g_{a\bc} + g_{a\bd}
 g_{b\bc} \label{eq:Fiibb1} \\
 \CF_{\ba b|c\bd} &=& g_{b\ba} g_{c\bd} \\
 \CF'_{ab|cd} &=& e^{\CK} \int_X \hat{\Omega} \wedge \partial_a \partial_b \partial_c
 \partial_d \hat{\Omega} \, \, \equiv \CF_{abcd}.
\end{eqnarray}
It should be noted that in general $D_a D_b D_c \Omega$ is
\emph{not} pure (1,3); there can be a (2,2)-part, since $\la D_a
D_b D_c \Omega,\bD_{\bd} \bD_{\be} \bOmega \ra = D_a
\CF_{bc|\bd\be} = D_a R_{b\bd c \be}$, which in general is
nonvanishing.

\subsection{Orientifold limit}

\noindent Things simplify considerably in orientifold limits of
the F-theory compactification. Then $X = (T^2 \times Y) /\IZ_2$
with $Y$ a Calabi-Yau threefold, which is equivalent to type IIb
on the corresponding orientifold of $Y$ with constant
dilaton-axion $\tau$.

There are $n=h^{2,1}_-(Y)$ complex structure moduli of $Y$
surviving the orientifold projection, and $2h^{2,1}_-(Y)+2$ fluxes
can be turned on \cite{BH,Grimm:2004uq}.\footnote{The minus sign
refers to the part of the cohomology odd under the orientifold
involution.} Let $z^0 = \tau$ and $z^i$ ($i=1,\ldots,n$) be the
complex structure moduli of $Y$. Then
\begin{equation}
 \hat{\Omega}_4=\hat{\Omega}_1(t^0) \wedge \hat{\Omega}_3(t^i),
\end{equation}
so $\CK=\CK_1 + \CK_3$ with $\CK_1=-\ln(i \la
\hat{\Omega}_1,\hat{\bOmega}_1 \ra)=-\ln(2 \, \im \tau)$ and
$\CK_3=-\ln(i \la \hat{\Omega}_3,\hat{\bOmega}_3 \ra)$, and the
metric and curvature components mixing $0$ and $i$ all vanish. In 
general, there may be other fourfold moduli as well, which take $X$
away from the orientifold limit. These correspond to D7-brane moduli
from the IIb point of view. We will ignore them in what follows.

As before, we define normalized holomorphic forms by $\Omega_r =
e^{\CK_r/2} \hat{\Omega}_r$. Using the same methods as we used
before to compute intersection products of derivatives of
$\Omega_4$, one obtains
\begin{equation}
 D_0 \Omega_1 = \CF_0 \bOmega_1, \quad D_0 D_0 \Omega_1=0
\end{equation}
with $\CF_0 \equiv i \la \Omega_1,D_0 \Omega_1 \ra = i e^{\CK_1}
\la \hat{\Omega}_1,\partial_0 \hat{\Omega}_1 \ra =
-1/(\tau-\btau)$, and
\begin{equation}
 D_i D_j \Omega_3 = \CF_{ijk} \bD^k \bOmega_3,
\end{equation}
with $\CF_{ijk} \equiv i \la \Omega_3,D_i D_j D_k \Omega_3 \ra = i
e^{\CK_3} \la \hat{\Omega}_3,\partial_i \partial_j \partial_k
\hat{\Omega}_3 \ra$. Therefore all components of $\CF_{ab|\bc\bd}$
and $\CF_{abcd}$ are zero, except
\begin{eqnarray}
 \CF_{0i|\bar{0}\bj} &=& g_{0\bar{0}} \, g_{i\bj} \\
 \CF_{ij|\bk\bl} &=& {\CF_{ij}}^{\bm} \bar{\CF}_{\bm\bk\bl}
 \label{eq:Fiibb2} \\
 \CF_{0ijk} &=& \CF_0 \CF_{ijk}.
\end{eqnarray}
Note also that $D_i D_j \Omega_4 = \CF_{0ijk} \bD^0 \bD^k
\bOmega_4$.

The space of allowed fluxes $H^4_F(X)$ consists of harmonic
4-forms $G_4 = -\alpha \wedge F_3 + \beta \wedge H_3$, where
$\{\alpha,\beta\}$ is a canonical basis of harmonic 1-forms on
$T^2$ (such that $\hat{\Omega}_1=\beta - \tau \alpha$) and $F_3$
and $H_3$ are harmonic 3-forms on $Y$ (identified with type IIb
R-R resp.\ NS-NS flux). The main simplification occurs because
$\Omega$, $D_a\Omega$, $D_0 D_i \Omega$ and their complex
conjugates form a Hodge-decomposition basis of $H_F^4(X)$. To see
this, note that we have $\dim H_F(X) = 2(2n+2)$, which equals the
number of vectors in the proposed basis set, and that linear
independence of this set follows from the intersection products
computed earlier. The basis can be turned into an orthonormal
basis by introducing an orthonormal frame $e^a_A$ for the metric
on moduli space, $\delta_{A\bB}=e^a_A g_{a\bb} e^{\bb}_{\bB}$,
where capital letters refer to the frame indices.\footnote{For
explicit numerical indices $0,1,\ldots$ we will underline frame
indices, but only if confusion could arise.} We take
$e^{\un}_0=\CF_0$, so $\CF_{\un}=1$. The basis $\CB=\{\Omega,D_A
\Omega,D_{\un} D_I \Omega \} \cup \{c.c.\}$ now satisfies
\begin{equation}
 \la \CB,\bar{\CB}\ra=\mbox{diag}(1,-{\bf 1}_{n+1},{\bf 1}_n,1,-{\bf
1}_{n+1},{\bf 1}_n).
\end{equation}
Various physical quantities have a simple expression in terms of
components with respect to this basis. Writing
\begin{equation} \label{eq:Hodgebasis}
 G_4 = \bX \Omega - \bY^A D_A \Omega + \bZ^I D_{\un} D_I \Omega +
 c.c.,
\end{equation}
we get for example for the flux superpotential \eq{fluxW} and
its derivatives (transformed to the orthonormal frame by $D_A
\cdots D_B \equiv e^a_A \cdots e^b_B \, D_a \cdots D_b$):
\begin{eqnarray}
 W &=& \la G_4,\Omega \ra = X \label{eq:WXYZ} \\
 D_A W &=& \la G_4,D_A \Omega \ra = Y_A \label{eq:DWXYZ} \\
 D_{\un} D_{\un} W &=& 0 \\
 D_{\un} D_I W &=& Z_I \\
 D_I D_J W &=& \CF_{IJK} \bZ^K \label{eq:DDWXYZ} \\
 D_{\un} D_I D_J W &=& \CF_{IJK} \bY^K \\
 D_I D_J D_K W &=& (D_I \CF_{JKL}) \bZ^L + \CF_{IJK} \bY^0,
 \label{eq:DDDWXYZ}
\end{eqnarray}
for the potential
\begin{equation}
 V = |Y|^2-3|X|^2,
\end{equation}
and for the flux induced D3-charge tadpole
\begin{equation}
 L=\frac{1}{2} N \eta N = \frac{1}{2}\la G_4,G_4 \ra = |X|^2 - |Y|^2 + |Z|^2.
\end{equation}

\section{Distributions of supersymmetric vacua}

A supersymmetric flux vacuum is characterized by a choice of $K$
flux quanta $N^\alpha$ and a solution to $D W = 0$. We wish to
compute the total number of such flux vacua satisfying the
constraint $L \equiv \frac{1}{2} N \eta N \leq L_*$,
\begin{eqnarray}
 \CN_{susy}(L \leq L_*) &=& \sum_{\mbox{susy vac}} \theta(L-L_*) \\
 &=& \frac{1}{2 \pi i} \int_C \frac{d\alpha}{\alpha} e^{\alpha
 L_*} \CN(\alpha) \label{eq:laplacetransform}
\end{eqnarray}
where $C$ runs along the imaginary axis passing zero to the right,
and where we introduced the Laplace transformed ``weighted
number'' of vacua
\begin{eqnarray}
 \CN(\alpha) &\equiv& \sum_{vac} e^{- \frac{\alpha}{2} N \eta N} \\
 &=& \sum_N \int_{\CM} d^{2m} z \, \delta^{2m}(DW) \,
 |\det D^2W| \\
 &\approx& \int_{\CM} d^{2m} z \int d^K N
 \, e^{- \frac{\alpha}{2} N \eta N} \, \delta^{2m}(DW) \,
 |\det D^2W|. \label{eq:intapprox}
\end{eqnarray}
In the last step we approximated the sum over fluxes by an
integral. By rescaling $N \to N/\sqrt{\alpha}$, it is easy to see
that \eq{intapprox} scales simply as $\alpha^{-K/2}$, so in this
approximation \eq{laplacetransform} gives:
\begin{equation}
 \CN_{susy}(L\leq L_*) = \theta(L_*) \frac{{L_*}^{K/2}}{(K/2)!} \, \CN(\alpha=1).
\end{equation}

As discussed in the previous section, in the orientifold limit we
have $m=n+1$ with $n$ the number of complex structure moduli of
$Y$, and $K=4 m$. In this case, it is possible to directly
evaluate the Gaussian integral by changing variables from $N$ to
$(X,Y,Z,\bX,\bY,\bZ)$, related to each other by the Hodge
decomposition \eq{Hodgebasis}:
\begin{equation}
 N = \eta^{-1} (\bX \Pi - \bY^A D_A \Pi + \bZ^I D_{0} D_I \Pi +
 c.c.).
\end{equation}
The Jacobian for this change of variables is
\begin{equation}
 J = 2^{2m} |\det M| = 4^m |\det \eta|^{-1/2} |\det(M^{\dagger} \eta
 M)|^{1/2},
\end{equation}
where $M=\eta^{-1}(\Pi,-D_A \Pi,D_{0}D_I\Pi,c.c.)$. The extra
factor $2^{2m}$ accounts for the fact that for complex variables
we use the convention $d^2 z =\frac{1}{2i} dz \wedge d\bz$.
Happily, because of the orthonormality of our Hodge decomposition
basis $\CB$, we have
\begin{equation}
 M^{\dagger} \eta M = \mbox{diag}(1,-{\bf 1}_{n+1},{\bf 1}_n,1,-{\bf
1}_{n+1},{\bf 1}_n),
\end{equation}
hence the Jacobian is simply $J = 4^m |\det \eta|^{-1/2}$.
Furthermore, from \eq{DWXYZ}, we get
\begin{equation}
 \delta^{2m}(D_a W)=|\det e^A_a|^{-2} \delta^{2m}(D_A W)= (\det
 g)^{-1} \delta^{2m}(Y_A)
\end{equation}
and from \eq{WXYZ}--\eq{DDWXYZ} together with $D_a \bD_{\bb}
W = g_{a\bb} W$:
\begin{eqnarray}
 (\det g)^{-2} \det D^2W &=&
 \det [(D_{0},D_I,\bD_{0},\bD_I)^t \cdot (\bD_{0}\bW,\bD_J \bW,D_{0}W,D_J
 W)] \\
 &=&\det \left(
 \begin{array}{cccc}
   \bX & 0 & 0 & Z_J \\
   0 & \delta_{IJ} \bX & Z_I & \CF_{IJK} \bZ^K \\
   0 & \bZ_J & X & 0 \\
   \bZ_I & \bar{\CF}_{IJK} Z^K & 0 & \delta_{IJ} X
 \end{array}
 \right) \label{eq:origdet} \\
 &=&
  \det \left(
 \begin{array}{cccc}
   \bX & 0 & 0 & Z_J \\
   0 & X & \bZ_J & 0 \\
   0 & Z_I & \delta_{IJ} \bX & \CF_{IJK} \bZ^K \\
   \bZ_I & 0 & \bar{\CF}_{IJK} Z^K & \delta_{IJ} X
 \end{array}
 \right) \label{eq:origdet2} \\
 &=& |X|^2 \det \left(
  \begin{array}{cc}
    \delta_{IJ} \bX - \frac{Z_I \bZ_J}{X} & \CF_{IJK} \bZ^K \\
    \bar{\CF}_{IJK} Z^K & \delta_{IJ} X - \frac{\bZ_I Z_J}{\bX}.
  \end{array}
 \right) \label{eq:detDDW}
\end{eqnarray}
Putting everything together, we find for the total number of
supersymmetric vacua:
\begin{equation} \label{eq:rhodef}
 \CN(L \leq L_*) = \frac{(2 \pi L_*)^{2m}}{(2m)!} |\det \eta|^{-1/2}
 \int_{\CM} d^{2m}\!z \det g \, \rho(z)
\end{equation}
where
\begin{equation} \label{eq:rho}
  \rho(z) = \pi^{-2m} \int d^2 X d^{2n} Z \,
  e^{-|X|^2-|Z|^2} \,
 |X|^2 |\det \left(
  \begin{array}{cc}
    \delta_{IJ} \bX - \frac{Z_I \bZ_J}{X} & \CF_{IJK} \bZ^K \\
    \bar{\CF}_{IJK} Z^K & \delta_{IJ} X - \frac{\bZ_I Z_J}{\bX}.
  \end{array}
 \right)|.
\end{equation}
The function $\rho$ measures the density of supersymmetric vacua
per unit volume in moduli space. It is specified entirely in terms
of the special geometry data $\CF_{IJK}$. In particular, $\rho$
has no dependence on the dilaton modulus $\tau$, and therefore the
integration over the fundamental $\tau$-domain $\CM_{\tau}$ in
\eq{rhodef} simply contributes a factor $\mbox{vol}(\CM_{\tau})
= \pi/12$.

Similarly to $\rho$, we define the index density $\rho_{ind}$,
counting vacua with signs, by dropping the absolute value signs
from the determinant in \eq{rho}.

\subsection{Computing densities}

\subsubsection{The case $n=1$}

\noindent The total susy vacuum number density for $n=1$ can be
computed explicitly from \eq{rho}:
\begin{eqnarray}
 \rho &=& \pi^{-4} \int d^2 X d^2 Z \,
  e^{-|X|^2-|Z|^2} \, ||X|^4+|Z|^4-(2+|\CF|^2)|X|^2|Z|^2| \\
  &=& \pi^{-2} \int dr ds \, e^{-r-s} \, |r^2+s^2-(2+|\CF|^2) r
  s| \label{eq:rhoint} \\
  &=& \pi^{-2}(2 - |\CF|^2 + \frac{2 |\CF|^3}{\sqrt{4+|\CF|^2}}).
\end{eqnarray}
This can be obtained by splitting up the integration domain in
three parts, separated by the lines $s/r=\frac{1}{2}(2+|\CF|^2 \pm
|\CF|\sqrt{4+|\CF|^2})$ on which the determinant changes sign. The
first two terms in this expression correspond to the index
density:
\begin{equation}
 \rho_{ind}=(2-|\CF|^2)/\pi^2.
\end{equation}
In the large complex structure limit one has universally
$|\CF|=2/\sqrt{3}$ and, in the special coordinate $t$,
$g_{t\bt}=-3/(t-\bar{t})^2$ (to verify the former, recall that
$\CF=(e^t_{\underline{1}})^3 \CF_{ttt} = (g_{t\bt})^{-3/2}
e^{\CK_3} i \hat{\Pi} \eta^{-1}
\partial_t^3 \hat{\Pi}$, with $e^{\CK_3}=-\frac{i}{k (t-\bt)^3}$ and
$\hat{\Pi} \eta^{-1} \partial_t^3 \hat{\Pi} = 6 k$). So
$\rho_{LCS} = 2/\pi^2$ (and $\rho_{ind}=\rho/3$), and if we
approximate the large complex structure region $\CM_{LCS}$ by the
standard fundamental domain in the upper half plane, we get
\begin{equation} \label{eq:lcscontribution}
 \int_{\CM_{\tau}} \!\! d^2\tau g_{\tau\bar{\tau}} \int_{\CM_{LCS}}\!\! d^2 t
 g_{t\bt} \, \rho \approx \frac{1}{12} \times \frac{3}{12}
 \times 2 = \frac{1}{24}.
\end{equation}

Near a conifold point (or more generally a discriminant locus),
$\CF$ blows up (see example 2 below). Note that in that case, up
to a sign, the total number density equals the index density. In
fact this is true for any $m$: if all $\CF_{IJK} \to \infty$, the
terms involving $\CF_{IJK}$ in \eq{rho} will dominate the
determinant, so
\begin{equation} \label{eq:conifrho}
 \det D^2 W \approx (-1)^n |X|^2 |\det \CF_{IJK} \bZ^K|^2
\end{equation}
and putting the absolute value signs around $D^2 W$ only removes
the overall $(-1)^n$.

\subsubsection{Index density}

Computing the total density for $n>1$ becomes hard. However, the
index density can be given a simple expression in terms of
geometric quantities \cite{AD}. One way to do this is to rewrite
the determinant as a Gaussian over Grassmann variables, and then
to perform first the Gaussian over $X$ and $Z$, and next the the
Grassmann integral. Using $R_{I\bar{J}K\bar{L}}={\CF_{IK}}^M
\bar{\CF}_{MJL}-\delta_{IJ}\delta_{KL}-\delta_{IL}\delta_{JK}$
(which follows e.g.\ from comparing \eq{Fiibb1} with \eq{Fiibb2})
and $R_{\un\un\un\un}=-2$, one gets after some manipulations
\begin{equation} \label{eq:indexformula}
 d\mu_{ind} = d^{2m}z \det g \, \rho_{ind}  = \pi^{-m}
 \det(R+\omega {\bf 1}),
\end{equation}
with $R$ the curvature form and $\omega$ the K\"ahler form on
$\CM$. This is in agreement with earlier results \cite{AD}. Note
however that the index density does \emph{not} factorize,
$\det(R+\omega {\bf 1})_{T^2 \times Y} \neq \det(R+\omega {\bf
1})_{T^2} \wedge \det(R+\omega {\bf 1})_{Y}$, so the claim in v1
of \cite{AD} that adding the dilaton to the moduli just multiplies
the index by $1/12$ was not correct. For example for $n=1$, one
has
\begin{eqnarray}
 d\mu_{ind}&=&\pi^{-2} \det \left( \begin{array}{cc} R_0 + \omega_0 + \omega_1 & 0 \\
 0 & R_1 + \omega_0 + \omega_1 \end{array} \right) \\
 &=& \pi^{-2} (-\omega_0+\omega_1) \wedge (R_1+\omega_0+\omega_1) \\
 &=& - \pi^{-2} \omega_0 \wedge R_1.
\end{eqnarray}
In this case, we therefore simply have
\begin{equation} \label{eq:indn1}
 \int_{\CM} d\mu_{ind} = - \frac{1}{12} \chi(\CM_Y).
\end{equation}

\subsubsection{Example 1: $T^6$}

\noindent As a toy example, let us take $Y$ to be the
$(T^2)^3/\IZ_2$ orientifold with the $T^6$ and the fluxes
restricted to be diagonal and symmetric under permuations of the
three $T^2$'s. Then the complex structure moduli space is the
fundamental domain in the upper half plane, parametrizing the
$T^2$ modulus, and \eq{lcscontribution} is exact. The orientifold
has 64 O3-planes, so $L_*=16$. A basis for the symmetric fluxes is
$\{\Sigma_{\alpha}\}_{\alpha=1,\cdots,4}$, with the
$\Sigma_{\alpha}$ given by the generating function
$\sum_{\alpha=1}^4 \Sigma_{\alpha} t^{\alpha-1} = \prod_{k=1}^3
(\alpha_k+t \, \beta_k)$. Here $(\alpha_k,\beta_k)$ is a canonical
basis of $H^1(\IZ)$ of the $k$th $T^2$. The nonvanishing
intersection products on $T^6$ are $\la \Sigma_1,\Sigma_4\ra=1$
and $\la \Sigma_2,\Sigma_3 \ra=3$. To avoid a subtlety with flux
quantization involving discrete fluxes on the $O3$ planes
\cite{Frey:2002hf}, we will furthermore as in \cite{Kachru:2002he}
restrict to even fluxes, i.e.\ we take as basis $\{ 2
\Sigma_\alpha \}$. The corresponding intersection form on $Y$ thus
has as nonzero entries $\eta^Y_{14}=-\eta^Y_{14}=2$ and
$\eta^Y_{23}=-\eta^Y_{32}=6$. The intersection form $\eta$ on $T^2
\times Y$ is the direct product of this with the $T^2$
intersection form $\epsilon_{ij}$. Therefore $|\det \eta|=(2^4
\times 9)^2$, and the total number of these flux vacua is,
according to \eq{rhodef}:
\begin{equation}
 \CN_{susy} = \frac{(2\pi \times 16)^4}{4!} \times (2^4 \times 9)^{-1} \times \frac{1}{24}
 = 1231.
\end{equation}

\subsubsection{Example 2: Conifold}

\noindent Let $Y$ be a Calabi-Yau manifold near a generic conifold
degeneration. For simplicity we only consider one modulus, namely
the period of the vanishing cycle $v = \int_A \hat{\Omega}$.
Inclusion of more moduli does not change the essential features.
The monodromy around $v=0$ implies that the period of the dual
cycle is of the form $\int_B \hat{\Omega} = - \frac{\ln v}{2 \pi
i} \, v \, + $ analytic terms. The metric near $v=0$ is then
\begin{equation}
 g_{v\bv} \approx c \, \ln \frac{\mu^2}{|v|^2}
\end{equation}
where $\mu$ is some constant and $c=e^{\CK_0}/2 \pi$ with $\CK_0$
the K\"ahler potential at $v=0$. Furthermore
\begin{equation} \label{eq:Fconif}
 \CF = g_{v\bv}^{-3/2} \, e^{\CK} \, \biggl( 
 i \mbox{$\int_A \hat{\Omega} \,
 \, \partial_v^3 \int_B \hat{\Omega}$} \, + \, \mbox{anal. } \biggr)
 \approx i \, \biggl( c \, \ln \frac{\mu^2}{|v|^2} \biggr)^{-3/2} \,
 \frac{c}{v},
\end{equation}
so as announced earlier, we see that $\CF \to \infty$ when $v \to
0$. The same is true for $\rho \approx |\CF|^2/\pi^2$. However,
the density integrated over the fundamental $\tau$-domain and
$|z|<R$ remains finite. For small $R$:
\begin{equation} \label{eq:NvacConif}
 \int d^2\tau g_{\tau\bar{\tau}}
 \int d^2 v \, g_{v\bv} \, \rho \approx \frac{1}{12 \ln
 \frac{\mu^2}{R^2}}.
\end{equation}
Note that the constant $c$ has dropped out of this expression.
Plugging this in \eq{rhodef}, we get for the number of susy vacua
with $L \leq L_*$ and $|v| \leq R$:
\begin{equation} \label{eq:NvacConif2}
 \CN_{vac} = \frac{\pi^4 L_*^4}{18 \ln
 \frac{\mu^2}{R^2}}.
\end{equation}
The logarithmic dependence on $R$ implies that a substantial
fraction of vacua are extremely close to the conifold point. For
example when $L_*=100$ and $\mu=1$, there are still about one
million susy vacua with $|v|<10^{-100}$. Interestingly, vacua very
close to conifold degenerations are precisely the desired ones in
the context of phenomenological model building, as they provide a
natural mechanism for generating large scale hierarchies
\cite{Giddings:2001yu}, and may enable controlled constructions of
de Sitter vacua by adding anti-D3 branes, as proposed by KKLT
\cite{KKLT}. However, for the latter it is also necessary that the
mass matrix at the critical point is positive, and as we will see
below, this condition dramatically reduces the number of candidate
vacua.

\subsubsection{Example 3: Mirror Quintic}

The mirror quintic is given by a quotient of the hypersurface
$x_1^5 + x_2^5 + x_3^5 + x_4^5 + x_5^5 +  = 5 \, \psi \, x_1 x_2
x_3 x_4 x_5$ in $\IC P^4$. It has one complex structure modulus,
$\psi$, whose fundamental domain $\CM_Y$ is the wedge $-\pi/5<\arg
\psi< \pi / 5$. Its periods are well known \cite{Candelas} and can
be expressed as Meijer G-functions, which makes it possible to
study this case numerically.

\EPSFIGURE{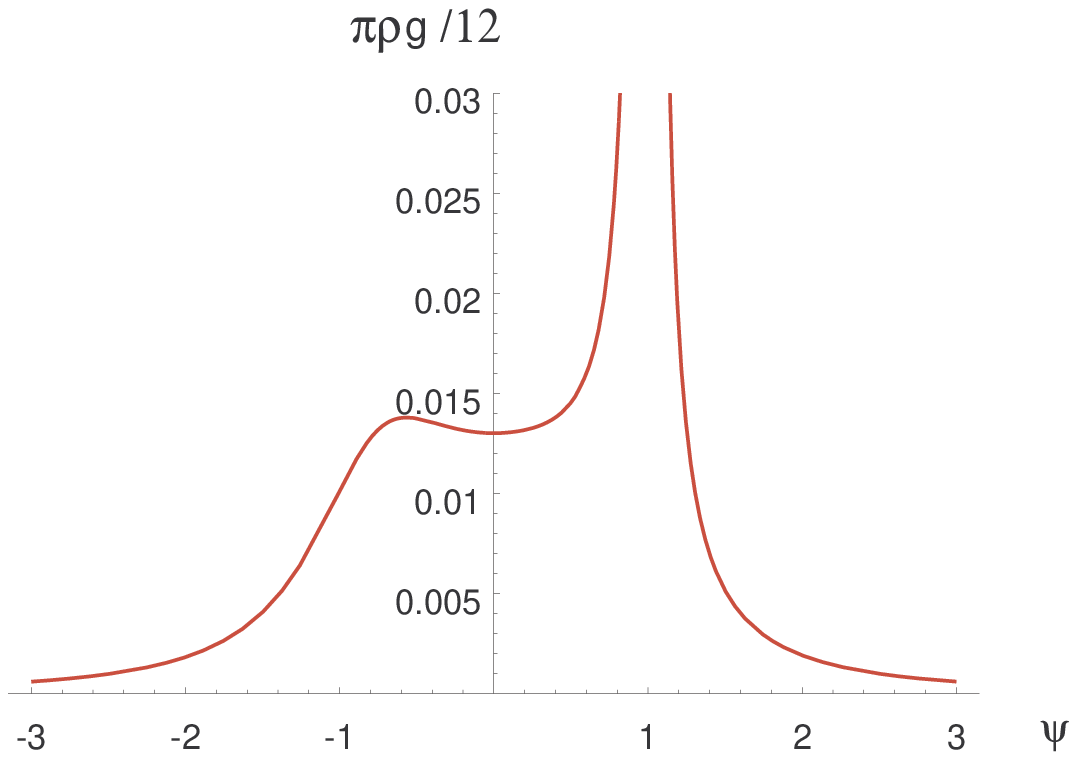,height=7cm,angle=0,trim=0 0 0 0}%
{The susy vacuum number density per unit
$\psi$ coordinate volume, $\pi \rho g_{\psi \bar{\psi}}/12$, on
the real $\psi$-axis, for the mirror quintic.
  \label{quinticsusydens}}

Fig.\ \ref{quinticsusydens} shows a plot of $\frac{\pi}{12} \,
\rho g_{\psi \bar{\psi}}$, i.e.\ the susy vacuum number density
per unit $\psi$ coordinate volume, on the real $\psi$-axis (the
factor $\pi/12$ comes from integrating over $\tau$). The drop for
$|\psi|>1$ is due to a similar drop in $g_{\psi \bar{\psi}}$;
$\rho$ itself tends to the large complex structure value $2/\pi^2$
when $\psi \to \infty$. The divergence at $\psi=1$ is due to the
presence of a conifold singularity there. In the notation of
example 2, the parameters specifying $g$ and $\CF$ near the
conifold are $\mu=8.94$ and $c=1.26 \times 10^{-2}$ (with $v
\approx -4 \pi^2 5^{-3/2} \, (\psi-1)$).

We numerically computed\footnote{This was done as follows. First,
we divided the moduli space in patches, since different regions
have different suitable coordinates and special care is required
near singularities. In each patch, the periods and their
derivatives were evaluated on a dense grid of points, and an
approximation of these functions was constructed by interpolation
(because direct evaluation of Meijer functions is much too
time-consuming). Finally from this data the various desired
quantities were constructed and integrated.} the integrated susy
vacuum number density. We found:
\begin{equation} \label{eq:intrho}
 \int_{\CM} d \mu = 5.46 \times 10^{-2}.
\end{equation}
This can be compared with an estimate of the large complex
structure contribution, obtained similar to
\eq{lcscontribution} by using the LCS expressions for $g$ and
$\rho$, but now cutting off the integral say at $\im t = 2$. (Here
$t$ defined by $5 \psi \equiv e^{-2 \pi i t/5}$, and the conifold
point is located at $\im t|_{\psi=1} = 2 \pi / 5 \ln 5 = 1.28$.)
The result is $1/ 16 \pi = 1.99 \times 10^{-2}$. The exact
numerical result for this region is almost the same: $1.97 \times
10^{-2}$. Thus for orientifolds of the mirror quintic, about 36\%
of all susy flux vacua are at $\im t > 2$ (and this fraction is
proportional to one over the lower bound on $\im t$). On the other
hand, using \eq{NvacConif}, we get that the fraction of vacua
with $|\psi-1|<S \ll 1$ equals $0.486/\ln(6.41/S^2)$. For
$S=10^{-3}$, this is about 3 \%, and for $S=10^{-10}$ still 1\%.

For the integrated index density, we found
\begin{equation}
 \int d\mu_{ind} = - 1.666 \times 10^{-2} \approx - 1/60
\end{equation}
Combined with \eq{indn1}, this indicates that $\chi(\CM_Y) =
1/5$. Indeed, this can be verified analytically. The integral of
the Euler class can be written as a sum of boundary contour
integrals as follows:
\begin{eqnarray}
 \chi &=& \frac{i}{2 \pi} \int_{\CM_Y} \bpartial \partial \ln
 g_{\psi\bpsi} \\
 &=& \frac{i}{2 \pi \times 5} \int_{\IC} \bpartial \partial \ln
 g_{\psi\bpsi} \\
 &=& \frac{i}{10 \pi} \biggl( \oint_{\infty} \partial \ln g_{\psi\bpsi}
  - \sum_{i=1}^5 \oint_{P_i} \partial \ln g_{\psi\bpsi} \biggr)
  \label{eq:eulercontours}
\end{eqnarray}
where the $P_i$ are the 5 copies of the conifold point in the
$\psi$-plane. For $\psi \to \infty$ we have $g_{\psi\bpsi} =
3/4|\psi|^2 \ln^2|\psi|$, so
\begin{equation}
 \partial \ln g_{\psi\bpsi} = -\frac{1}{\psi} \biggl( 1+\frac{1}{\ln |\psi|}
 \biggr) d\psi
\end{equation}
and the corresponding contour integral produces a contribution
$(i/10 \pi)\times(-2 \pi i) = 1/5$ to \eq{eulercontours}. On
the other hand, near the conifold point $\psi=1$, $g_{\psi\bpsi}=c
\ln(\mu^2/|\psi-1|^2)$, and
\begin{equation}
 \partial \ln g_{\psi \bpsi} = - \frac{1}{\ln
 \frac{\hat{\mu}^2}{|\psi-1|^2}} \frac{1}{\psi-1} \, d\psi,
\end{equation}
so the corresponding contour integral is zero. Adding up all
contributions, we thus see that $\chi = 1/5$.

Numerical integration of the volume of $\CM$ gives $5
\mbox{vol}(\CM)=3.1416 \approx \pi$. Again, this can be understood
topologically, using
\begin{equation}
 \omega = \frac{i}{2} \partial \bpartial \CK = \frac{i}{2} \bpartial \partial \ln i
 \la \hat{\Omega} , \bar{\hat{\Omega}} \ra,
\end{equation}
plus the fact that if $\hat{\Omega}$ is normalized such that $\CK$
is regular at $\psi=0$, we have $\la \hat{\Omega} ,
\bar{\hat{\Omega}} \ra \sim 1/|\psi|^2 \ln^3 |\psi|$ for $\psi \to
\infty$. Writing the volume integral as a sum over contours, again
only the $\psi = \infty$ contour contributes, and this
contribution equals $\pi$, as expected.

\subsubsection{$d^2|W|$ signature distribution}

\noindent For some applications, such as the RG flow
interpretation of domain wall flows, one needs to know whether the
critical point of $W$ is a maximum, a minimum, or a saddle point
of $|W|$. This corresponds to a positive, negative or indefinite
Hessian $d^2|W| \equiv (\partial_a,\bpartial_{\ba})^t \cdot
(\bpartial_{\bb} |W|,\partial_b|W|)$. At a critical point, one has
$\partial_a
\partial_b |W|=\frac{1}{2} \frac{\bW}{|W|} D_a D_b W$ and
$\partial_a \bpartial_b |W| = \frac{1}{2} \frac{W}{|W|} D_a
\bD_{\bb} \bW$, so from \eq{origdet} we see that we have to
investigate the eigenvalues of
\begin{equation} \label{eq:ddW}
 d^2|W|=\frac{1}{2 |X|} \left(
 \begin{array}{cccc}
   |X|^2 & 0 & 0 & \bX Z \\
   0 & |X|^2 & \bX Z & \CF \bX \bZ \\
   0 & X \bZ & |X|^2 & 0 \\
   X \bZ & \bar{\CF} X Z & 0 & |X|^2
 \end{array}
 \right)
\end{equation}
Clearly the sum of the eigenvalues $\tr d^2 |W| \geq 0$, so there
are no maxima (this is true in general, since $\partial_a
\bpartial_{\bb} |W| \sim g_{a\bb} |W|$ and $\tr g > 0$). In
general, a matrix is positive definite iff all upper left
submatrices have positive determinant. In the case at hand, this
implies the conditions $|X|^4+|Z|^4-(2+|\CF|^2) |X|^2 |Z|^2
>0$ and $|X|^2-|Z|^2>0$. This restricts the integration domain of
\eq{rhoint} to one of its three segments, and thus the density
of susy vacua which are minima of $|W|$ is
\begin{equation}
 \rho_{++++} = \frac{1}{2 \pi^2}(2 - |\CF|^2 + \frac{
 |\CF|^3}{\sqrt{4+|\CF|^2}}).
\end{equation}
More information can be obtained by looking directly at the
eigenvalues. We just quote the results: $\rho_{++--}=\rho_{++++}$,
and
\begin{equation}
 \rho_{+++-} = \frac{1}{\pi^2} \frac{
 |\CF|^3}{\sqrt{4+|\CF|^2}}.
\end{equation}
In the large complex structure limit one has, rather
democratically, $\rho_{++++}=\rho_{+++-}=\rho_{++--}=\rho/3$. In
the conifold limit on the other hand, $\rho_{+++-}=\rho$ and
$\rho_{++++}=\rho_{++--}=0$.

\subsection{Number of susy vacua with positive bosonic mass matrix}

Due to the properties of AdS, supersymmetric vacua are always
perturbatively stable, even if the critical point of the potential
$V$ is not a minimum. Obviously, this is no longer true if
supersymmetry is broken and the cosmological constant is lifted to
a positive value. In particular, if as in the KKLT scenario
supersymmetry is broken by adding an anti-D3 brane to a
supersymmetric AdS vacuum, thus shifting the potential up by a
constant\footnote{constant in the complex structure deformation
directions} such that the critical point of the potential becomes
positive, the original AdS critical point should be a minimum in
order for the lifted vacuum to be perturbatively stable. It is
therefore important to compute the number of supersymmetric vacua
which are local minima of $V$.

A critical point is a minimum if the mass matrix $d^2 V \equiv
(\partial,\bpartial)^t \cdot (\bpartial V,\partial V)$ is positive
definite. Using \eq{DV}--\eq{DcDV} and
\eq{WXYZ}--\eq{DDDWXYZ}, we get, at a supersymmetric critical
point (i.e.\ $Y=0$):
\begin{eqnarray} \label{eq:DDVsusy1}
 D_{0} \partial_{0} V &=& 0 \\
 D_I \partial_{0} V &=& - Z_I \bX \\
 D_{0} \partial_J V &=& - Z_J \bX  \\
 D_I \partial_J V &=&  - \CF_{IJK} \bZ^K \bX \\
 D_{0} \bpartial_{0} V &=& -2|X|^2+|Z_I|^2 \\
 D_I \bpartial_{0} V &=& \CF_{IKL} \bZ^K \bZ^L \\
 D_{0} \bpartial_J V &=& \bar{\CF}_{JKL} Z^K Z^L \\
 D_I \bpartial_J V &=& -2|X|^2 + Z_I \bZ_J
 + {\CF_{IK}}^M \bar{\CF}_{MJL} \bZ^K Z^L \label{eq:DDVsusy8}
\end{eqnarray}
We can use covariant derivatives of $V$ here instead of ordinary
derivatives because $d V=0$ implies $D d V =d^2 V$.

Let us work out the case $n=1$, which has
\begin{equation}
d^2 V= \left(
 \begin{array}{cccc}
   |Z|^2-2|X|^2 & \bar{\CF} Z^2 & 0 & -\bX Z \\
   \CF \bar{Z}^2 & (1+|\CF|^2)|Z|^2 -2|X|^2 & -\bX Z & -\CF \bX
   \bZ \\
   0 & - X \bZ & |Z|^2-2|X|^2 & \CF \bZ^2 \\
   -X \bZ & -\bar{\CF} X Z & \bar{\CF} Z^2 & (1+|\CF|^2)|Z|^2 -2|X|^2
 \end{array}
 \right)
\end{equation}
A matrix is positive definite iff all its upper left
subdeterminants are positive. With $r \equiv |X|^2$, $s \equiv
|Z|^2$, this gives the following conditions:
\begin{eqnarray}
 s-2 r &>& 0 \\
 4\,r^2 - 2\,\left( 2 + |\CF|^2 \right) \,r\,s + s^2 &>& 0\\
 \left( s - 2\,r \right) \,\left( 4\,r^2 - \left( 5 +
 2\,|\CF|^2 \right) \,r\,s + s^2 \right) &>&0 \\
 \left( 16\,r^2 - 4\,\left( 2 + |\CF|^2 \right) \,r\,s + s^2 \right) \,
   \left( r^2 - \left( 2 + |\CF|^2 \right) \,r\,s + s^2 \right)
   &>&0.
\end{eqnarray}
A straightforward analysis of these inequalities shows that they
boil down to simply
\begin{equation} \label{eq:posMcond}
 |Z|^2/|X|^2 = s/r > 4+2|\CF|^2+2|\CF|\sqrt{4+|\CF|^2}.
\end{equation}
To compute the number density of susy vacua with positive mass
matrix, we should therefore evaluate the integral \eq{rhoint}
with $(r,s)$ restricted to the region satisfying this condition.
This gives
\begin{equation} \label{eq:susyposMrho}
 \rho_{M>0} = \frac{98 + 179\,|\CF|^2 + 42\,|\CF|^4 + 96\,|\CF|\,{\sqrt{4 + |\CF|^2}} +
  42\,|\CF|^3\,{\sqrt{4 + |\CF|^2}}}
  {{\left( 5 + 2\,|\CF|^2 + 2\,|\CF|\,{\sqrt{4 + |\CF|^2}} \right) }^3\,{\pi }^2}
\end{equation}
as shown in fig.\ \ref{susydensities}. At $\CF=0$, the relative
fraction of susy vacua with positive mass matrix is approximately
39\% and maximal. In the large complex structure limit
$\CF=2/\sqrt{3}$ the fraction is about 19\%, and in the conifold
limit it is zero. Indeed, while the total density $\rho$ grows
quadratically with $|\CF|$ when $\CF \to \infty$, the density
$\rho_{M>0}$ \emph{decreases} quadratically with $|\CF|$. This
means that very near a conifold point, though there are many susy
vacua, only an extremely small fraction has positive mass matrix!

\EPSFIGURE{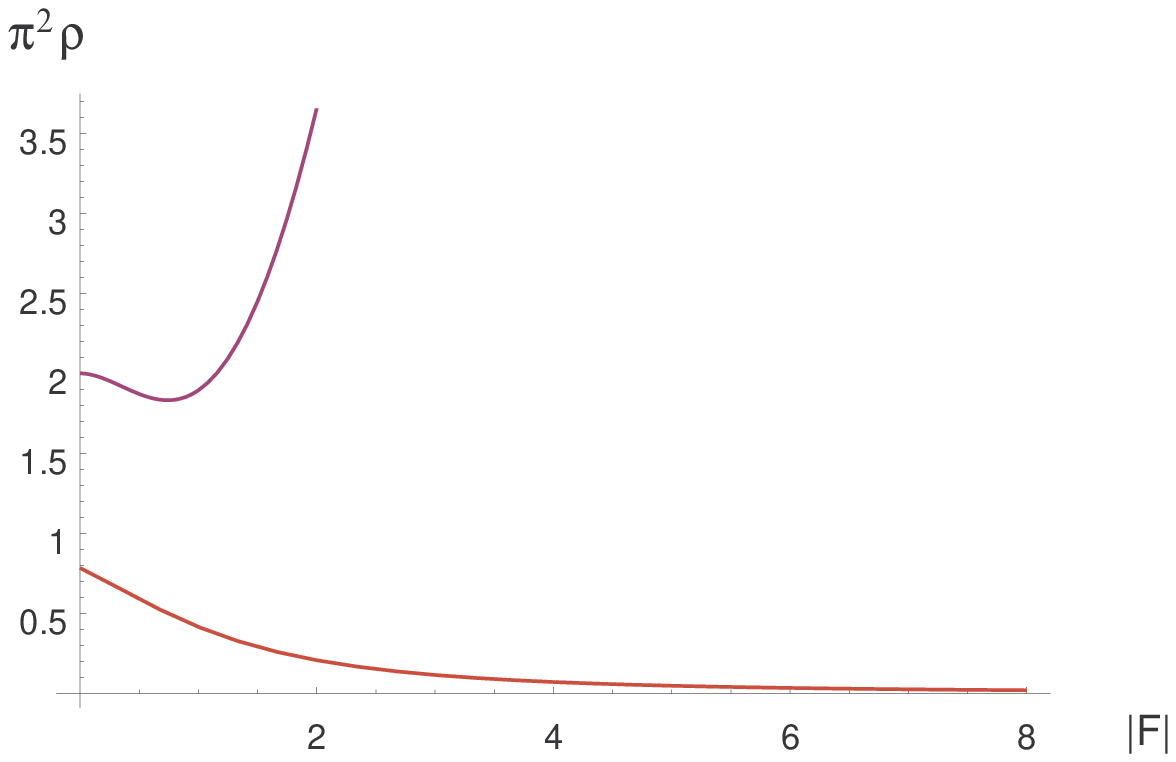,height=7cm,angle=0,trim=0 0 0 0} {The
upper curve shows the total susy vacuum density as a function of
$|\CF|$, the lower curve the density of susy vacua with positive
mass matrix. \label{susydensities}}

Let us make this more precise. In the notation of our example 2 in
the previous section, we have for small $v$:
\begin{equation}
 \rho_{M>0} \approx \frac{21 \, c^2}{6 \pi^2} \biggl(\ln \frac{\mu^2}{|v|^2}
 \biggr)^4 \, |v|^2.
\end{equation}
Integrated over $\tau$ and $|v|<R$, this gives
\begin{equation} \label{eq:NstablevacConif}
 \int d^2\tau g_{\tau\bar{\tau}}
 \int d^2 v \, g_{v\bv} \, \rho_{M>0}
 = \frac{7 c^2}{256} (3 + 6 \gamma + 6 \gamma^2 + 4 \gamma^3 + 2 \gamma^4) R^4,
\end{equation}
where $\gamma(R)=\ln \frac{\mu^2}{R^2}$. Because of the $R^4$
dependence, this rapidly goes to zero with $R$. If we take the
parameters of the mirror quintic conifold for example, i.e.
$\mu=8.94$ and $c=1.26 \times 10^{-2}$, and we take $L_* = 100$,
then the expected number of susy vacua with $M>0$ drops below 1
for $|v|<0.02$, or $|\psi-1|<0.004$. Increasing $L_*$ to 1000,
these numbers become just one order of magnitude smaller. Thus,
for $n=1$ and with reasonable parameter values, at most only a
modest hierarchy of scales can be generated through the conifold
throat mechanism of \cite{Giddings:2001yu}, if we insist on having
a positive mass matrix.

On the other hand, \emph{if} a near-conifold vacuum has positive
mass matrix, \eq{posMcond} together with $|X|^2+|Z|^2=L \leq L_*$
shows that it automatically has a small value for $|W|=|X|$, and
therefore a small cosmological constant. We will make this more
precise in the next section.

The positivity properties of $M=d^2 V$ for susy vacua can also be
analyzed as follows. First observe that in general, if $DW=0$,
\begin{equation} 
 M = H^2 - 3 |W| H,
\end{equation}
where
\begin{equation}
 H = 2 \, d^2 |W|.
\end{equation}
This follows directly from \eq{DDV}--\eq{DcDV}. Thus, to have
$M>0$, all eigenvalues $\lambda$ of $H$ must satisfy $\lambda < 0$
or $\lambda > 3 |W|$. In particular, if $W=0$ at the critical
point, $M$ is automatically non-negative, and by continuity the
same will be true for most susy vacua with small $W$. The
suppression of vacua with $M>0$ near a conifold point (for $n=1$)
can now be seen in the following way. According to \eq{ddW}, the
matrix $H$ can be written as $H=|X|\, {\bf 1} + \Delta H$, with
\begin{equation} \label{eq:Hmass}
 \Delta H = \left( \begin{array}{cc} 0 & S \\ \bar{S} & 0 \end{array}
 \right), \qquad
 S = \frac{\bX}{|X|} \left( \begin{array}{cc} 0 & Z \\ Z & \CF
 \bZ.
 \end{array} \right)
\end{equation}
The eigenvalues of $\Delta H$ are $(\pm \lambda_1,\pm \lambda_2)$,
with $\lambda_1^2 \lambda_2^2 = \det \Delta H = |Z|^4$ and
$2(\lambda_1^2 + \lambda_2^2) = \tr (\Delta H)^2 =
2(|\CF|^2+2)|Z|^2$. When $\CF \to \infty$, the eigenvalues of
$\Delta H$ are therefore approximately given by $\pm |\CF|^{\pm 1}
|Z|$, and the eigenvalues of $H$ by $\lambda = |X|\pm |\CF|^{\pm
1} |Z|$. The condition on $\lambda$ to have $M>0$ translates to
$|Z|>2 |\CF| |X|$, in agreement with what we found earlier.

Note that the ``dangerous'' eigenvectors (the eigenvectors of
$\Delta H$ with eigenvalues $\sim 1/|\CF|$) are approximately
aligned with the dilaton direction, i.e.\
$(1,O[1/\CF],1,O[1/\CF])$.  In a sense, the special form of the
matrix \eq{Hmass} leads to a sort of ``seesaw'' mixing with the
dilaton, and the small eigenvalue.  This may be specific to $n=1$;
a similar analysis for $n>1$ suugests that there is no longer
suppression of $M>0$ vacua near generic points of the discriminant
locus.

\EPSFIGURE{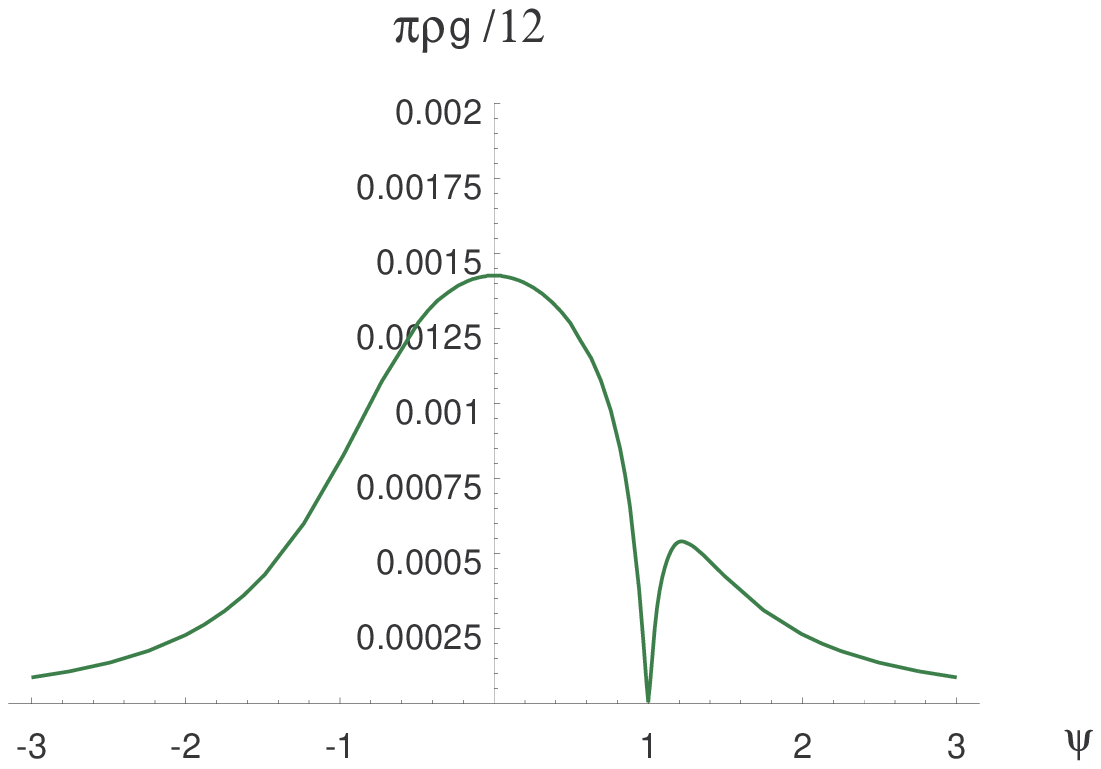,height=7cm,angle=0,trim=0 0 0
0} {The number density of susy vacua with positive mass matrix,
per unit coordinate volume, on the real
$\psi$-axis, for the mirror quintic. Compare to fig. %
\ref{quinticsusydens}.
\label{quinticsusyposMdens}}

As an example, a plot of the number density of $M>0$ vacua per
unit coordinate volume is shown in fig.\ \ref{quinticsusyposMdens}
for the mirror quintic, on the real $\psi$-axis. The sharp dip
near $\psi=1$ is due to the conifold singularity. For the
integrated density we find
\begin{equation}
 \int_{\CM} d\mu_{M>0} = 5.12 \times 10^{-3}.
\end{equation}
Comparing this to \eq{intrho}, we thus see that about 9\% of
all susy vacua has positive mass matrix. This is not too far from
the naive guess $1/16$ based on the fact that there are four mass
eigenvalues, with each a 50\% chance to be positive.

\subsection{Distribution of cosmological
constants}

The cosmological constant in a supersymmetric vacuum is $V=-3
|W|^2$ (in $2 T_3 \equiv 1$ units). We wish to count the number of
susy vacua with $L \leq L_*$ and $V \geq V_* \equiv -3 \lambda_*$,
i.e.\ with $|W|^2 \leq \lambda_*$. Analogous to
\eq{laplacetransform}\footnote{We could also have implemented the
$|W|^2 \leq \lambda_*$ inequality by a similar Laplace transform,
but here it is slightly easier to solve the constraints
directly.}, we have
\begin{equation} \label{eq:NvacCC}
 \CN_{susy}(L \leq L_*,|W|^2 \leq \lambda_*) =
 \int_{\CM} d^{2m}\!z \int_0^{\lambda_*} \! d\lambda \,
 \frac{1}{2 \pi i} \int \frac{d\alpha}{\alpha}
 e^{\alpha L_*} \nu(z,\alpha,\lambda)
\end{equation}
where
\begin{eqnarray}
 \nu(z,\alpha,\lambda) &=&
 \int d^K\! N \, e^{-\alpha L} \, \delta(|W|^2-\lambda) \, \delta^{2m}(DW) \,
 |\det D^2W|\\
 &=& \alpha^{1-K/2} \int d^K\! N \, e^{- L} \, \delta(|W|^2-\alpha \lambda) \,
 \delta^{2m}(DW) \, |\det D^2W|. \label{eq:rescconstr}
\end{eqnarray}
Parallel to \eq{rhodef}, we can write \eq{NvacCC} also in
the form
\begin{equation} \label{eq:NvacCC2}
 \CN_{susy}(L \leq L_*,|W|^2 \leq \lambda_*) =
 \frac{(2 \pi L_*)^{K/2}}{(K/2)!} |\det \eta|^{-1/2}
 \int_{\CM} d^{2m}\!z \, g \int_0^{\lambda_*} \! d\lambda \,
 \rho(\lambda,z)
\end{equation}
The quantity $d\lambda \, \rho(\lambda,z)/\rho(z)$ then gives the
fraction of vacua at $z$ with $|W|^2$ in a width $d\lambda$
interval around $\lambda$.

Specializing again to the orientifold limit, so $K=4m=4(n+1)$, we
make the change of variables from $N$ to $(X,Y,Z)$. In these
variables the additional constraint in \eq{rescconstr} is simply
$|X|^2=\alpha \lambda$, which we can solve by putting
$X=\sqrt{\alpha \lambda}$, since the integral is invariant under
an overall phase transformation of $(X,Y,Z)$. Thus we find,
similar to \eq{rho} (but with the additional constraint):
\begin{eqnarray}
 \nu(z,\alpha,\lambda)&=&\frac{2^{2m} \pi g}{\sqrt{\eta} \, \alpha^{2m-1}}
 \int d^{2n}\!Z \, e^{-\alpha \lambda-|Z|^2}
   \,
 |\det \left(
 \begin{array}{cccc}
   \sqrt{\alpha \lambda} & 0 & 0 & Z_J \\
   0 & \delta_{IJ} \sqrt{\alpha \lambda} & Z_I & \CF_{IJK} \bZ^K \\
   0 & \bZ_J & \sqrt{\alpha \lambda} & 0 \\
   \bZ_I & \bar{\CF}_{IJK} Z^K & 0 & \delta_{IJ} \sqrt{\alpha \lambda}
 \end{array}
 \right)\!| \nonumber \\
 &=& \frac{2^{2m} \pi g}{\sqrt{\eta} \, \alpha^{2m-1}} \, e^{-\alpha \lambda} \int d^{2n}\! Z \,
 e^{-|Z|^2}|\sum_{k=0}^m C_k(Z,\bZ) (\alpha \lambda)^k|
 \label{eq:absval}
\end{eqnarray}
where $g\equiv \det g$, $\eta=|\det \eta|$ and the $C_k(Z,\bZ)$
are homogeneous polynomial functions obtained by expanding the
determinant. For example in the $n=1$ case,
\begin{equation} \label{eq:fintn1}
\nu(z,\alpha,\lambda)=\frac{16 \pi g}{\sqrt{\eta} \, \alpha^3}
 \, e^{-\alpha \lambda} \int d^2\! Z \,
 e^{-|Z|^2} \, |(\alpha \lambda)^2-(2+|\CF|^2)|Z|^2 \alpha \lambda +
|Z|^4|.
\end{equation}

\subsubsection{Density at zero cosmological constant}

For physical applications, the most interesting quantity is the
number of vacua at very small cosmological constant, i.e.\ the
case $\lambda_* \ll L_*$. To a good approximation, we can then
neglect the higher order terms in \eq{absval} and simply compute
the density at $\lambda=0$. The integral becomes
\begin{equation} \label{eq:IF}
 I(\CF) = \int d^{2n}\!Z \, e^{-|Z|^2} |
 \det \left(
 \begin{array}{cc}
  0 & Z_J \\ Z_I & \CF_{IJK} \bZ^K
 \end{array}
 \right)
 |^2.
\end{equation}
This is a significant simplification compared to \eq{absval}, as
this can be evaluated using Wick's theorem or by rewriting the
determinant as a Grassmann integral and then doing the integral
over $Z$, similar to what was done to obtain \eq{indexformula}.
Doing the integral over $\alpha$ in \eq{NvacCC} now gives
\begin{equation}
 \rho|_{\lambda=0} = \frac{2 \pi m}{\pi^{2m} L_*} I(\CF)
\end{equation}
Note that for $n=1$, this is independent of $\CF$ and therefore of
$z$: $\rho_0 = 8 / \pi^2 L_*$. Applying this to the example of the
mirror quintic, for which $\mbox{vol}(\CM_\tau \times \CM_{\psi})
= \frac{\pi}{12} \times \frac{\pi}{5}$, we get for small
$\lambda_*$:
\begin{equation} \label{eq:smallestCC}
 \CN(L\leq L_*,|W|^2 \leq \lambda_*) = \frac{4 \pi^4
 L_*^4}{45} \, \frac{\lambda_*}{L_*}
\end{equation}
For $L_*=100$, this becomes $\sim 10^7 \lambda_*$, so the expected
smallest possible cosmological constant is $|\Lambda| \sim 10^{-7}
T_3$. For $L_*=1000$, this is three orders of magnitude smaller.

\EPSFIGURE{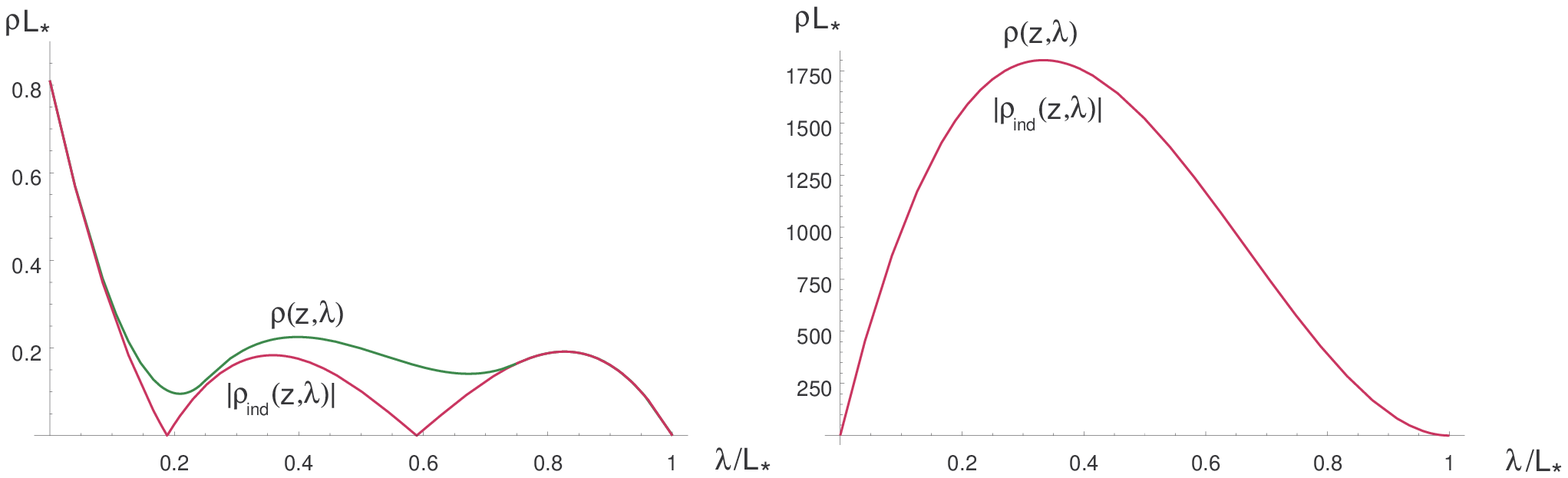,height=5.1cm,angle=0,trim=0 0 0 0}
{\emph{Left:} Vacuum number densities (true and absolute value of
index) in the large complex structure limit ($\CF=2/\sqrt{3}$), as
a function of cosmological constant value. \emph{Right:} Same near
the conifold limit ($\CF=100$ for this example). Note that
despite appearances, $\rho$ is non-zero at $\lambda=0$ as
expressed in \eq{smallestCC}; it just does not receive the
conifold enhancement there.
\label{CCdistr}}

\subsubsection{Index density}

The distribution for arbitrary cosmological constants is harder to
compute, mainly because of the absolute value signs in
\eq{absval}. Let us therefore drop these for now, so we count
vacua with signs. Then after integration over $Z$, we will still
have a polynomial in $\alpha$:
\begin{equation}
 \nu_{ind}(z,\alpha,\lambda) = \frac{2^{2m} \pi^m g}{\sqrt{\eta}} \, e^{-\alpha \lambda}
 \sum_{k=0}^m c_k \lambda^k \alpha^{k-2m+1}
\end{equation}
and therefore, by comparing \eq{NvacCC} and \eq{NvacCC2},
\begin{eqnarray} \label{eq:rhoindexpansion}
 \rho_{ind}(z,\lambda) &=& \frac{(2m)!}{(2 \pi L_*)^{2m}g}
 \frac{1}{2 \pi i} \int \frac{d\alpha}{\alpha} e^{\alpha L_*}
 \nu_{ind}(z,\alpha,\lambda) \\
 &=&  \theta(L_*-\lambda) \frac{(2m)!}{\pi^{m}L_*}
 \sum_{k=0}^m \frac{c_k}{(2m-1-k)!} \left(\frac{\lambda}{L_*}\right)^k
 \left(1-\frac{\lambda}{L_*}\right)^{2m-1-k}
\end{eqnarray}
In the case $n=1$, we thus get for the cosmological constant
density counted with signs
\begin{equation}
 \rho_{ind}(z,\lambda) =  \frac{4}{L_* \pi^2} (1 - x)
 ( 2 - ( 10 + 3 |\CF|^2)\,x  +
 (14 + 3 |\CF|^2 )\, x^2) \, \theta(1-x)
\end{equation}
where we wrote $x \equiv \frac{\lambda}{L_*}$. As noted above, the
density at $\lambda=0$ equals $8/L_* \pi^2$. In the large complex
structure limit $\CF \to 2/\sqrt{3}$, $\rho_{ind} = 8
(1-x)(1-7x+9x^2)/L_* \pi^2$ and in the conifold limit $\CF \to
\infty$, for $x$ not too small, $\rho_{ind}/|\CF|^2 = -12(1-x)^2
x/L_* \pi^2$.

\subsubsection{Total density}

Let us now turn to the true density. The absolute value signs in
\eq{absval} make the integral hard to evaluate in general, but for
$n=1$ one gets something of the form
\begin{equation}
 \nu(z,\alpha,\lambda) = P_0(\alpha \lambda) e^{-\alpha \lambda} +
 P_-(\alpha \lambda) e^{-b_- \alpha \lambda}
 + P_+(\alpha \lambda) e^{-b_+ \alpha \lambda}
\end{equation}
with the $P_i$ polynomials and $b_-$, $b_+$ some $\CF$-dependent
coefficients (see below). Doing the $\alpha$-contour integral,
this then gives, explicitly:
\begin{equation}
 L_* \pi^2 \, \rho(z,\lambda) = L_* \pi^2 \rho_{ind}(z,\lambda) + P(x) (\theta(1-b_- x)-\theta(1- b_+ x))
 + Q(x) (\theta(1-b_- x) + \theta(1- b_+ x))
\end{equation}
where
\begin{eqnarray}
 P(x)&=&-4+6(4+|\CF|^2)\,x-12(4+|\CF|^2)\,x^2 +
 (32-6|\CF|^4-|\CF|^6)\,x^3 \\
 Q(x)&=& (4+|\CF|^2)^{3/2} F^3 x^3 \\
 b_{\pm} &=& \frac{1}{2}\sqrt{4+|\CF|^2}(\sqrt{4+|\CF|^2} \pm |\CF|)
\end{eqnarray}
In the large complex structure limit
\begin{equation}
 L_* \pi^2 \rho(z,\lambda) = L_* \pi^2 \rho_{ind}(z,\lambda) +
 16(1-4x)^2 \, \theta(1-4x)
 +\frac{16}{27} (16x-3)(3-4 x)^2 \, \theta(3-4x),
\end{equation}
and in the conifold limit, for $x$ not too small,
\begin{equation}
 \rho(z,\lambda)/|\CF|^2 \to  12(1-x)^2 x / L_* \pi^2 =
 -\rho_{ind}(\lambda,z).
\end{equation}
This is not immediately obvious from the above expressions, but
can be seen from \eq{fintn1}: the middle term dominates, and the
absolute value just removes the minus sign. As discussed before,
the convergence of index density and true density can be expected
to hold more generally near degenerations where $\CF_{IJK} \to
\infty$ (see also below).

These considerations are illustrated for the large complex
structure  and conifold limits in fig.\ \ref{CCdistr}. Note that
the approximation of the true density by the index density is
perfect near the conifold, and near the extremities of $\lambda$,
and qualitatively still not bad away from those.

\subsubsection{Some general observations}

A few simple observations can be made about the general case. It
is obvious that the cosmological constant always satisfies
$|W|^2<L_*$. This is because $|W|^2+|Z|^2 = L \leq L_*$ by
construction. The density at $\lambda=L_*$ is zero for the same
reason; in fact, the index density near that point is suppressed
by $(L-\lambda_*)^{m-1}$, as follows from \eq{rhoindexpansion}. By
contrast, the density at $\lambda=0$ is always nonvanishing. Note
however that in case all $\CF_{IJK} \to \infty$, this value is
much smaller than the total density $\rho(z)$ (compare \eq{IF} to
\eq{conifrho}). In general, for large $m$, the distribution can be
expected to peak at a small value of $\lambda$, because of the
suppression factor $(L-\lambda_*)^{m-1}$. More intuitively, the
lower $|W|^2$ is, the more ``room'' there is for $Z$ to satisfy
$|W|^2+|Z|^2 \leq L_*$.

Finally, in the limit where $\CF_{IJK} \to \infty$, we can compute
$\rho(z,\lambda)$ more explicitly, since then, for $\lambda$ not
too small,
\begin{eqnarray}
 \rho(z,\lambda) &\approx& \frac{(2m)!}{(\pi L_*)^{2m}}
 \frac{1}{2 \pi i} \int \frac{d\alpha}{\alpha} e^{\alpha L_*}
  \frac{\pi}{\alpha^{2m-1}} \int d^{2n}\!Z \, e^{-\alpha \lambda-|Z|^2}
   \, \alpha \lambda |\det (\CF_{IJK} \bZ^K)|^2 \\
   &=& A \, x \, (1-x)^{2m-2}
\end{eqnarray}
where $x=\lambda/L_*$, and $A = \frac{2m(2m-1)}{\pi^{2m-1} L_*}
\int d^{2n}\!Z \, e^{-|Z|^2} \, |\det (\CF_{IJK} \bZ^K)|^2$. This
distribution has mean $\mu=\la x \ra=2/(2m+1)$ and standard
deviation $\sigma^2=\la (x-\mu)^2 \ra = (2m-1)/(m+1)(2m+1)^2$, so
for large $m$, most susy vacua in this limit have a cosmological
constant $\Lambda$ of order $-T_3 L_*/m$. This should be compared
to the string scale energy density, which in Einstein frame is
$T_F^2=g_s (2\pi\alpha')^{-2} = 2 \pi g_s T_3$. Therefore in this
limit vacua with $|\Lambda|$ well below the string scale are
comparatively rare, unless $m \gg L_*/g_s$.

\subsubsection{Distribution restricted to vacua with positive mass
matrix}

If we restrict to vacua with positive mass matrix,
\eq{posMcond} together with $|Z|^2+|X|^2 = L \leq L_*$, implies
the cutoff $\lambda \leq L_*/(5+2|\CF|^2+2|\CF|\sqrt{4+|\CF|^2})$.
In the large complex structure limit, this is $\lambda \leq
L_*/13$, and near the conifold limit, $\lambda \leq L_*/4
|\CF|^2$. Therefore, positive $M$ vacua near the conifold point
automatically have small cosmological constant (small compared to
the string scale). Recall however that positive $M$ vacua are
suppressed near the conifold point. We saw earlier that for the
parameter values of the mirror quintic and $L_* = 100$, no
positive $M$ vacua are expected below $v \sim 10^{-2}$. At this
point $\CF \sim 20$ and thus $\lambda < 10^{-3} L_* = 0.1$, hence
the cosmological constant cutoff is not much below the string
scale. Increasing $L_*$ to 1000 decreases the $\lambda$ cutoff
with just one order of magnitude. Thus, for $n=1$, the positive
$M$ vacua closest to the conifold point are not expected to be
close enough to force the cosmological constant to be
hierarchically smaller than the string scale.

This does not mean of course that there are no $M>0$ vacua with
very small cosmological constant; in fact, the vacuum density at
$\lambda_* = 0$ is the same for $M>0$-vacua as for vacua without
constraints on $M$, since if $X=0$, the $M>0$ condition
\eq{posMcond} is automatically satisfied. Most vacua with very
small $\lambda$ will therefore have a positive mass matrix. This
is actually true for any supergravity theory, as follows from
\eq{MHH}.

\subsection{Counting attractor points}

The techniques we used to count supersymmetric flux vacua can also
be used to count supersymmetric black holes, or attractor points,
in type IIB theory compactified on a Calabi-Yau $X$. More
precisely, we wish to count the number of duality inequivalent,
regular, spherically symmetric, BPS black holes with entropy $S$
less than $S_*$. The charge of a black hole is given by an element
$Q = N^\alpha \Sigma_\alpha$ of $H^3(X,\IZ)$. The central charge
$\CZ= \la Q,\Omega_3 \ra$ plays a role similar to the (normalized)
superpotential $W$: the moduli at the horizon are always fixed at
a critical point of $|\CZ|$, i.e.\ $D \CZ = (\partial + \partial
\CK) \CZ = 0$. The entropy is then given by $S = A/4 = \pi
|\CZ|^2$, evaluated at the critical point. For a given $Q$,
critical points may or may not exist, may or may not be located in
the fundamental domain $\CM$, and may or may not be unique, but
clearly we can label all equivalence classes of black holes by a
charge vector $N$ and a critical point within the fundamental
domain. Thus our desired number is, in a continuum approximation
(valid for large $S_*$) similar to what we had before:
\begin{equation}
 \CN_{BH}(S \leq S_*) = \frac{1}{2 \pi i} \int \frac{d\alpha}{\alpha}
 \, e^{\alpha S_*/\pi} \int d^K\!N \int_{\CM} d^{2n}\!z \, e^{-\alpha
 |\CZ|^2} \, \delta^{2n}(D\CZ) \, |\det D^2 \CZ|.
\end{equation}
Here $K$ is the number of flux components and $n$ the number of
complex structure moduli, $K=2n+2$. After rescaling $N \to
N/\sqrt{\alpha}$, doing the $\alpha$ integral becomes
straightforward, and we get
\begin{equation}
 \CN_{BH}(S \leq S_*) = \frac{(2 S_*)^{n+1}}{(n+1)!} \int_{\CM} d^{2n}\!z \,
 \det g \, \rho(z)
\end{equation}
where
\begin{equation} \label{eq:attrrho1}
 \rho(z) = \frac{1}{(2 \pi)^{n+1} \det g} \int d^{2n+2}\!N \, e^{-|\CZ|^2}
 \delta^{2n}(D\CZ) \, |\det D^2 \CZ|.
\end{equation}
To evaluate this integral, we change variables to a
Hodge-decomposition basis of $H^3(X)$, similar to
\eq{Hodgebasis}:
\begin{equation}
 Q = i \bar{X} \Omega - i \bar{Y}^I D_I \Omega + c.c.,
\end{equation}
so $\CZ = \la Q,\Omega \ra = X$ and $D_I \CZ = \la Q, D_I \CZ \ra
= Y_I$. Capital indices again refer to an orthonormal frame. The
Jacobian for the change of variables from $N$ to $(X,Y,\bX,\bY)$
can be computed in a way analogous to what we did for flux vacua;
the result is $J = 2^{n+1} |\eta|^{-1/2}$, where $|\eta|$ is the
determinant of the intersection form (equal to 1 if we sum over
the full charge lattice). Furthermore, using the special geometry
identity $D_I D_J \CZ = \CF_{IJK} Y^K$, we get that at a critical
point (where $Y=0$):
\begin{equation}
 \det \left(
 \begin{array}{cc}
  D_I \bD_J \bar{\CZ} & D_I D_J \CZ \\
  \bD_I \bD_J \bar{\CZ} & \bD_I D_J \CZ
 \end{array}
 \right)
 =\det \left(
 \begin{array}{cc}
  \delta_{IJ} \bar{X} & 0 \\
  0 & \delta_{IJ} X
 \end{array}
 \right)
 = |X|^{2n}.
\end{equation}
Plugging this in \eq{attrrho1} gives
\begin{eqnarray}
 \rho &=& \frac{1}{\pi^n |\eta|^{1/2}} \int d^2\!X \int d^{2n}\!Y
 \,e^{-|X|^2} \, \delta^{2 n}(Y) \, |X|^{2n} \\
 &=&\frac{n!}{\pi^n |\eta|^{1/2}}.
\end{eqnarray}
The factor $\det g$ dropped out because of the change from
coordinate to orthonormal frame. Note that this expression for
$\rho$ is independent of $z$. This means that attractor points are
uniformly distributed over moduli space. Our final result is, for
large $S_*$:
\begin{equation} \label{eq:BHcount}
 \CN_{BH}(S \leq S_*) = \frac{2^{n+1} \, \mbox{vol}(\CM)}{(n+1)\,
 \pi^n \, |\eta|^{1/2}} \, {S_*}^{n+1}.
\end{equation}
For the mirror quintic, $n=1$ and $\mbox{vol}(\CM)=\pi/5$, hence
\begin{equation}
 \CN_{BH}(S \leq S_*) = \frac{2}{5} {S_*}^2
\end{equation}

The problem of counting the number of duality inequivalent black holes with given entropy was studied
in \cite{Moore:1998pn} using number theory techniques. In particular, for
$Y=T^2 \times K3$, this was shown to be related to class numbers and the number of inequivalent embeddings
of a given two-dimensional lattice into the charge lattice.

To compare to our results, we restrict to
algebraic K3's, as the complex structure moduli moduli space of a generic K3 is not Hausdorff, so the volume
factor in \eq{BHcount} would not make any sense. An algebraic K3 has moduli space 
$$
 \CM_{K3} = O(\Lambda)\setminus O(2,k) /O(2) \times O(k), 
$$
where $\Lambda$ is the charge lattice, which is the orthogonal complement of the Picard lattice and has 
signature $(2,k)$. This space is Hausdorff and has complex dimension $k$. A black hole charge is specified
by two charge vectors $p,q \in \Lambda$.

In this case, counting black holes
with given entropy, or more precisely with given ``discriminant'' $D=(S/\pi)^2$ (which is always an integer for 
$K3 \times T^2$), amounts roughly to computing the number of inequivalent lattice embeddings in $\Lambda$ of
lattices spanned by $q$ and $p$ with determinant $D$.\footnote{We thank Greg Moore for explaining this to us.}
This number can  in principle be obtained from the Smith-Minkowski ``mass formula'', which was used in
\cite{Moore:1998pn} to derive an estimate for the asymptotic growth of the number $N(D)$ of inequivalent black holes 
with discriminant $D$. The result is $N(D) \sim D^{k/2}$ and consequently 
the number of black holes with $D \leq D_*$ will grow as $D_*^{1+k/2}$.

To compare with \eq{BHcount}, note that $n=k+1$, so we get
\begin{equation}
 \CN_{BH}(D \leq D_*) = \frac{2^{k+2} \pi}{k+2} \mbox{vol}(\CM_{T^2}) \mbox{vol}(\CM_{K3}) \, D_*^{1 + k/2},
\end{equation}
which agrees with the growth given above, but is a bit more precise. Turning things around,
this formula should give a predicition for the asymptotic behavior of the Smith-Minkowski mass formula.

\section{Distributions of nonsupersymmetric vacua}

A general flux vacuum, supersymmetric or not, is characterized by
a flux vector and a critical point of the corresponding potential
$V$, which for now we do not require to be a local minimum. For
nonsupersymmetric vacua, the condition $L \leq L_*$ is no longer
sufficient to guarantee a finite volume of allowed fluxes.
Physically, it is reasonable to put also a bound on the
supersymmetry breaking parameter: $|DW|=|Y| \leq F_*$. Such a
bound, together with the bound on $L=|X|^2-|Y|^2+|Z|^2$, is indeed
precisely what we need to make the volume of allowed fluxes
finite.

In our approximation, the total number of flux vacua satisfying
these bounds is then given by
\begin{equation} \label{eq:nonsusyint}
 \CN(L \leq L_*,|DW| \leq F_*) = 2^{2m} \int d^{2m}\! z \int d^2\!X d^{2m}\!Y d^{2n}\!Z \,
 \theta(L_*-L)\,
 \theta(F_*-|Y|) \, \delta^{2m}(d V) \, |\det d^2 V|
\end{equation}
where $d^2 V=(\partial,\bpartial)^t \cdot (\bpartial V,\partial
V)$, and using \eq{DV}--\eq{DcDV} and
\eq{WXYZ}--\eq{DDDWXYZ}:
\begin{eqnarray}
 \partial_{0} V &=& Z_I \bY^I - 2 Y_{0} \bX \\
 \partial_I V &=& Z_I \bY^{0} + \CF_{IJK} \bY^J \bZ^K - 2 Y_I \bX \\
 D_{0} \partial_{0} V &=& 0 \\
 D_I \partial_{0} V &=& \CF_{IKL} \bY^K \bY^L - Z_I \bX \\
 D_{0} \partial_J V &=& \CF_{JKL} \bY^K \bY^L - Z_J \bX  \\
 D_I \partial_J V &=& 2 \CF_{IJK} \bY^{0} \bY^K - \CF_{IJK} \bZ^K
 \bX + D_I \CF_{JKL} \bZ^L \bY^K \\
 D_{0} \bpartial_{0} V &=& -2|X|^2-2|Y_{0}|^2+|Y_I|^2+|Z_I|^2 \\
 D_I \bpartial_{0} V &=& -Y_I \bY_{0} + \CF_{IKL} \bZ^K \bZ^L \\
 D_{0} \bpartial_J V &=& - Y_{0} \bY_J + \bar{\CF}_{JKL} Z^K Z^L \\
 D_I \bpartial_J V &=& -2|X|^2+|Y_{0}|^2 -2 Y_I \bY_J + Z_I \bZ_J
 + {\CF_{IK}}^M \bar{\CF}_{MJL}  (\bY^K Y^L + \bZ^K Z^L)
\end{eqnarray}
We can use covariant derivatives of $V$ in the integral instead of
ordinary derivatives because if $d V=0$, then $D d V =d^2 V$.

\subsection{Supersymmetric and anti-supersymmetric branches}

The main complication arises because the constraint $d V = 0$ is
quadratic. It defines a cone in $\IC^m$, which has several
branches. One obvious branch is $Y_A=0$, with $X$ and $Z_I$
arbitrary. This corresponds to the supersymmetric vacua discussed
in the previous sections. Another obvious one is $X=Z_I=0$. As we
will see, the vacua on this branch behave in a way as
``anti-supersymmetric'' vacua. For example, while susy vacua have
imaginary self-dual fluxes, these have imaginary anti-self-dual
fluxes. There are more complicated ``intermediate'' branches too,
which arise when the constraints considered as linear equations in
$(X,Z)$ (or in $Y$) are degenerate.

On the supersymmetric branch, we have $\delta^{2m}(d V)=|\det
M|^{-1} \delta^{2m}(Y)$, with $M=(\partial_Y,\bpartial_{\bY})^t
\cdot (\partial V,\bpartial V)$. This threatens to make the
integral divergent when $\det M \to 0$, but note that by the chain
rule and because $Y=0$, $d^2
V|_{Y=0}=(\partial,\bpartial)^t\cdot(\bpartial V,\partial V)=
(\partial,\bpartial)^t\cdot(\bY,Y)\cdot(\bpartial_{\bY},\partial_Y)^t\cdot(\bpartial
V,\partial V) = D^2 W \cdot \bar{M}$, so the factor $\det M$
cancels out of \eq{nonsusyint} and we are left with the
integral we had before for the supersymmetric case, as we should
of course.

Similarly, on the branch $X=Z=0$, we have $\delta^{2m}(d V)=|\det
M|^{-1} \delta^{2m}(X,Z)$, with
$M=(\partial_X,\partial_Z,\bpartial_{\bX},\bpartial_{\bZ})^t \cdot
(\partial V,\bpartial V)$. Again $\det M$ cancels out, because
$d^2 V|_{X=Z=0}=A \cdot \bar{M}$, with
\begin{eqnarray}
 A &=& (D_0,D_I,\bD_0,\bD_I)^t \cdot (\bW,\bD_0 \bD_J \bW,W,D_0 D_J W) \\
 &=&\left(
 \begin{array}{cccc}
  0 & \bY_J & Y_0 & 0 \\
  0 & \delta_{IJ} \bY_0 & Y_I & \CF_{IJK} \bY^K \\
  \bY_0 & 0 & 0 & Y_J \\
  \bY_I & \CF_{IJK} Y^K & 0 & \delta_{IJ} Y_0
 \end{array}
 \right),
\end{eqnarray}
so the remaining determinant factor in \eq{nonsusyint} is
\begin{equation}
 \det A = |Y_0|^2 \det \left(
 \begin{array}{cc}
  \delta_{IJ} \bY_0 - \frac{Y_I \bY_J}{Y_0} & \CF_{IJK} \bY^K \\
  \bar{\CF}_{IJK} Y^K & \delta_{IJ} Y_0 - \frac{\bY_I Y_J}{\bY_0}
 \end{array}
 \right),
\end{equation}
which is identical to the $\det D^2 W$ factor \eq{detDDW} of
the supersymmetric case, after substitution $X \to Y_0$, $Z_I \to
Y_I$. If we moreover replace the supersymmetric constraint $L =
|X|^2 + |Z|^2 \leq L_*$ by $|Y|^2 \leq L_*$ (or, since now
$L=-|Y|^2$, by $L \geq -L_*$), the integral for this
nonsupersymmetric branch is therefore formally the same as the one
for the supersymmetric branch! Thus
\begin{equation}
 \CN_{X=Z=0}(|DW| \leq F_*) = \CN_{susy}(L \leq F_*^2).
\end{equation}
In fact this can be understood directly from the defining
equations. If $(F_3,H_3,\tau,z^i)$ with $\im \tau>0$ solves $F_3 -
\tau H_3=i
* (F_3 - \tau H_3)$, which is the condition for a supersymmetric
vacuum, then $(F'_3,H'_3,\tau',z'^i)=(-F_3,H_3,-\bar{\tau},z^i)$
has $\im \tau'>0$ and solves $F'_3 - \tau'H'_3=-i*(F'_3 - \tau'
H'_3)$, which is the condition for a nonsupersymmetric vacuum on
the branch under consideration. An equivalent way to see this map
is to observe that $D_0 (N_{RR}+\tau N_{NS}) \sim -N_{RR}-
\bar{\tau} N_{NS}$ together with the fact that the action of $D_0$
and $\bD_0$ interchanges $(X,Z)$ and $Y$. Note however that this
is \emph{not} a map between vacua with the same topological data,
since it maps $L \to -L$; the susy vacua have positive and the
nonsusy vacua negative $L$.

The cosmological constant of these nonsupersymmetric vacua is
$\Lambda=2 T_3 |DW|^2=-2 L \, T_3$. Since $L$ is quantized, this
is always at least of the order of the string scale energy density
$T_F^2=2 \pi g_s T_3$, so actually the field theory approximation
on which this analysis is based cannot be trusted, and the
existence of these vacua in the full theory is doubtful.

\subsection{Intermediate branches}

The anti-supersymmetric branch $X=Z=0$ of the constraint cone $d V
= 0$ is parametrized by the values of the F-terms $Y_A$. For
generic values of the $Y_A$, the unique solution to $d V = 0$ is
indeed $X=Z=0$, since it is just a generic, complete system of
linear equations in $(X,Z)$. However, for some values of $Y$ the
linear system can become degenerate, namely when $\det M = 0$
where $M$ is as before given by
$M=(\partial_X,\partial_Z,\bpartial_{\bX},\bpartial_{\bZ})^t \cdot
(\partial V,\bpartial V)$. This happens at the intersection with
other branches.

We will only analyze the case $n=1$ here. Then
\begin{eqnarray}
 \partial_0 V &=& Z_1 \bY_1 - 2 Y_0 \bX \\
 \partial_1 V &=& Z_1 \bY_0 + \CF \, \bY_1 \bZ_1 - 2 Y_1 \bX
\end{eqnarray}
and $\det M = -4(|Y_0|^4 + |Y_1|^4 - (2+|\CF|^2) |Y_0|^2
|Y_1|^2)$. The equation $\det M = 0$ has two branches of
solutions:
\begin{equation}
 |Y_1|^2 = \lambda_{\pm}^2 |Y_0|^2, \qquad \mbox{with }
 \lambda_{\pm} = - \frac{1}{\lambda_{\mp}} = \frac{1}{2}(|\CF| \pm \sqrt{4+|\CF|^2}).
\end{equation}
These branches can be parametrized for example by $(Y_0,Z_1)$. The
explicit solutions of the constraints $d V=0$ are then, for
generic $(Y_0,Z_1)$:
\begin{eqnarray} \label{eq:explsol}
 Y_1 &=& \lambda_{\pm} \, e^{i(\arg F - 2 \arg Z_1)} \, Y_0 \\
 X &=& \frac{1}{2} \lambda_{\pm} \, e^{i(\arg F - 2 \arg Z_1 + 2 \arg Y_0)} \,
 \bZ_1.
\end{eqnarray}
Solving the constraint in this way, the delta-function $\delta^4(d
V)$ produces a factor $|\det \tM|^{-1}$ with $\tM =
(\partial_{Y_1},\partial_X,\bpartial_{\bY_1},\bpartial_{\bX})^t
\cdot (\partial V,\bpartial V)$. On both branches we have
\begin{equation}
\label{eq:detM}
 |\det \tM| = 4 |\CF| \sqrt{4 + |\CF|^2} \, |Y_0|^2 |Z_1|^2.
\end{equation}
Furthermore, in general,
\begin{equation}
 d^2V = \left(
 \begin{array}{cc}
   R & S \\
   \bar{S} & \bar{R}
 \end{array}
 \right),
\end{equation}
where for $n=1$
\begin{eqnarray}
 R &=& \left(
 \begin{array}{cc}
    |Z_1|^2 -2 |X|^2 - 2 |Y_0|^2 + |Y_1|^2 & \bar{\CF} Z_1^2 - Y_0 \bY_1 \\
   \CF \bZ_1^2 - \bY_0 Y_1 & |Z_1|^2 -2|X|^2 + |Y_0|^2 - 2 |Y_1|^2 +
   |\CF|^2(|Y_1|^2+|Z_1|^2)
 \end{array}
 \right) \nonumber \\
 S &=& \left(
 \begin{array}{cc}
 0 & \CF \, \bY_1^2 - \bX Z_1 \\
 \CF \, \bY_1^2 - \bX Z_1 & \CF (2 \bY_0 \bY_1 - \bX \bZ_1) +
 D_1 \CF \, \bY_1 \bZ_1
 \end{array}
 \right). \nonumber
\end{eqnarray}

\subsubsection{Large complex structure limit}

The general expression for $\det d^2 V$ is obviously very
complicated, but at large complex structure, where
$\CF=2/\sqrt{3}$ and $D \CF = 0$, things simplify considerably. In
this case, we get for the integration measure on the first branch,
with $r \equiv |Y_0|^2, s \equiv |Z_1|^2$:
\begin{equation}
 |\det d^2 V|/|\det \tM| = \frac{1}{12} |(12 r - 5 s)(8 r-3 s)|
\end{equation}
and on the second branch
\begin{equation}
 |\det d^2 V|/|\det \tM| = \frac{25}{972} |(24 r - 7
 s)(4 r+3 s)|.
\end{equation}
The flux number $L$, susy breaking parameter $|Y|$ and
cosmological constant $V$ for the first branch is (still in the
large complex structure limit)
\begin{eqnarray}
 L &=& -4 r + \frac{7 \, s}{4} \\
 |Y|^2 &=& 4 r \\
 V &=& 4 r - \frac{9\,s}{4}
\end{eqnarray}
and for the second one
\begin{eqnarray}
 L &=& -\frac{4\,r}{3} + \frac{13\,s}{12} \\
 |Y|^2 &=& \frac{4\,r}{3} \\
 V &=& \frac{4\,r}{3} - \frac{s}{4}
\end{eqnarray}
It is most convenient to implement the constraint on $L$ in
\eq{nonsusyint} by doing a Laplace transform, and the one on
$|Y|^2$ by constraining $r$ directly. The result for arbitrary
$F_*$ and $L_*$ is a bit complicated and not very instructive ---
it involves several polynomials multiplied by different
step-functions --- but for $L_*$ positive and $F_*$ not too large,
the step functions are unimportant, and we get a simple
polynomial. On branch 1, for $x \equiv F_*^2/L_* < 6$:
\begin{equation}
  \CN_{1,lcs}(L \leq L_*,|DW| \leq F_*) = x \, (3.07-0.499 \, x + 0.0256 \, x^2+1.04 \, x^3) \,
  V \, L_*^4
\end{equation}
where $V = \mbox{vol}(\CM_\tau \times \CM_{lcs}) = (\pi/12)
\mbox{vol}(\CM_{lcs})$. On branch 2, for $x<42/25$:
\begin{equation}
  \CN_{2,lcs}(L \leq L_*,|DW| \leq F_*) = x \, (5.59 + 0.416 \, x - 2.31 \, x^2+1.88 \, x^3) \,
  V \, L_*^4.
\end{equation}
This can be compared to the number of supersymmetric flux vacua at
large complex structure,
\begin{equation}
 \CN_{susy}(L \leq L_*) = \frac{(2 \pi L_*)^4}{24} \,
 \frac{2}{\pi^2} \, V = 13.2 \,
 V \, L_*^4.
\end{equation}
We still have to investigate if the vacua are perturbatively
stable. For vacua with positive cosmological constant, this is the
case if and only if $d^2 V$ is positive definite. A matrix is
positive iff all upper left subdeterminants are positive, which
here translates in the conditions
\begin{eqnarray}
 2 \, r - s &>& 0 \\
 -(12 \, r - 7 \, s)(4\,r - 3\,s) &>& 0 \\
 -(2\, r - s)(96 \,r^2 - 68\, r\, s + 15\, s^2) &>&0 \\
 r\, s\, (12 \,r - 5 \,s)(8\,r - 3\,s) &>&0
\end{eqnarray}
on branch 1 and similar conditions on branch 2. Straightforward analysis of these systems of
inequalities shows that they have no solutions, on either branch.
Alternatively, we could have computed the eigenvalues of $d^2V$.
For the first branch for example, these are
\begin{equation}
 \{ 4\, r, 4 \, r - \frac{5\,s}{3}, - \frac{2\,s}{3}, -8 \, r + 3\,
 s \}
\end{equation}
Again we see it is impossible to have only positive eigenvalues.
Thus we find, somewhat surprisingly, that there are no meta-stable
de Sitter vacua at large complex structure on these branches.

\subsubsection{Conifold limit}

\noindent In the conifold limit, we have, as in \eq{Fconif}:
\begin{equation} \label{eq:Fconifeps}
 \CF = \frac{i}{c^{1/2} \gamma^{3/2} v} \equiv \frac{1}{\epsilon}.
\end{equation}
where $\gamma=\ln \frac{\mu^2}{|v|^2}$. Now $D \CF$ is no longer
zero:
\begin{equation}
 D_{\underline{v}} \CF = \frac{\partial_v |\CF|^2}{g^{1/2} \bar{\CF}}
 = -\frac{i (1-3/\gamma)}{c \, \gamma^2 \, v^2}
 = \frac{i(\gamma-3)}{\epsilon^2}.
\end{equation}
However $D\CF$ turns out to drop out of all relevant quantities to
leading order in $\epsilon$.

From \eq{explsol}, it follows that on branch 1, to leading
order, $|Y_1|=|Y_0|/|\epsilon|$ and $|X|=|Z_1|/2|\epsilon|$, and
similarly but with opposite power of $\epsilon$ on branch 2. On
branch 1, we thus have, to leading order:
\begin{eqnarray}
 L &=& (|Z_1|^2/4 - |Y_0|^2)/|\epsilon|^{2} \equiv |z|^2/4 - |y|^2 \\
 |Y|^2 &=& |Y_0|^2/|\epsilon|^{2} = |y|^2 \\
 V &=& |y|^2 - 3 |z|^2/4
\end{eqnarray}
Since the constraints in \eq{nonsusyint} keep $L$ and $|Y|$
finite, $Z_1$ and $Y_0$ are of order $\epsilon$, and the rescaled
variables $y,z$ are of order 1. In these variables the integration
measure is, to leading order in $\epsilon$:
\begin{equation}
 d^2 Y_0 \, d^2 Z_1 \, \left | \frac{\det d^2 V}{\det \tM} \right|
 = d^2 y  \, d^2 z \, \frac{|y|^2 |z|^2}{16 |\epsilon|^2}.
\end{equation}
Doing the integral over $y$ and $z$ (implementing the constraints
as before) yields for \eq{nonsusyint}
\begin{equation}
 \CN_{1,con} = \frac{2 \pi^2}{3} \left( L_*^4 \theta(L_*) - (L_*-3 F_*^2)(L_*+F_*^2)^3
 \theta(L_* + F_*^2) \right)  \int d^2\tau \, d^2
 v  \frac{\det g}{|\epsilon|^2}.
\end{equation}
Apart from the $L_*$-dependent bracket, this is the same as the
expression for the number of near-conifold supersymmetric vacua,
so using \eq{NvacConif2} we immediately get for the number of
nonsusy critical points on branch 1 with $|v|<R$, $L<L_*$ and
$|Y|<F_*$:
\begin{equation}
\CN_{1,con} = \frac{\pi^4}{18 \ln \frac{\mu^2}{R^2}} \left( L_*^4
\theta(L_*) - (L_*-3 F_*^2)(L_*+F_*^2)^3
 \theta(L_* + F_*^2) \right)
\end{equation}
For $L_*$ positive, the $L_*$-dependent bracket reduces to $6
L_*^2 F_*^4 + 8 L_* F_*^6 +3 F_*^8$.

The same computation can be done for branch 2. To leading order in
$\epsilon$:
\begin{eqnarray}
 L &=& |Z_1|^2 - |Y_0|^2 \equiv |z|^2 - |y|^2 \\
 |Y|^2 &=& |y|^2 \\
 V &=& |y|^2 - 3 |\epsilon|^2 |z|^2/4 \label{eq:Vconif2} \\
 d^2 Y_0 \, d^2 Z_1 \, \left | \frac{\det d^2 V}{\det \tM} \right|
 &=& d^2 y  \, d^2 z \, \frac{|y|^2 |z|^2}{|\epsilon|^2}.
\end{eqnarray}
The final result is identical, $\CN_{1,con}=\CN_{2,con}$.

Do these near-conifold nonsusy critical points have positive mass
matrix? On the first branch, the first subdet condition is $2 \, r
- s >0$ (again with $r \equiv |y|^2$ and $s \equiv |z|^2$), and
using this, the third condition becomes
\begin{equation}
 (16 + O[\epsilon^2])\, r^2 +
 (2/|\epsilon|^2 + O[\epsilon^0]) \,r\,s +
 (1 + O[\epsilon^2] ) \, s^2 < 0.
\end{equation}
For small $\epsilon$, this is obviously false, so $d^2 V$ can
never be positive definite on this branch in the near-conifold
region; $\CN_{1,con,M>0}=0$. For the second branch on the other
hand, the positivity conditions are, after some simplifications
and dropping terms which are negligible\footnote{This includes
putting $\gamma + O[1] \approx \gamma$. It is possible to keep
terms of lower order in $\gamma$, but this only complicates the
formulas without changing the essential features.} for all $r$,
$s$ at sufficiently small $\epsilon$:
\begin{eqnarray}
 s/r &>& 4/|\epsilon|^2 \\
 -\gamma^2 \,s^2 + 4\,r\,s / |\epsilon|^2 +
 \gamma \sin \theta \,r^{-1/2} s^{5/2} |\epsilon| &>& 0,
\end{eqnarray}
where $\theta = \arg \CF + \arg Y_1 - \arg Z_1$ and $\gamma$ as
under \eq{Fconifeps}. With $s/r \equiv 4\, u^2 / |\epsilon|^2$,
this becomes
\begin{eqnarray}
 u - 1 &>& 0  \\
 p(u) \equiv \gamma \sin \theta \, u^3 - \gamma^2 \, u^2 + 1 &>& 0.
\end{eqnarray}
The polynomial $p(u)$ has three real roots $u_i$. We have
$p'(0)=0$, $p''(0)=-2\gamma^2<0$, $p(0)=1>0$ and $p(1) < 0$.
Therefore, if $\sin \theta < 0$, then $u_1<u_2<0<u_3<1$ and
$p(u)<0$ for $u>1$, so the system of inequalities has no
solutions. When $\sin \theta < 0$ on the other hand, we have
$u_1<0<u_2<1<u_3$, so the above system of inequalities boils down
to
\begin{equation}
 u>u_3 \approx \gamma / \sin \theta.
\end{equation}
The latter approximation becomes exact in the extreme conifold
limit $\gamma \to \infty$, but is already very good for $\gamma >
4$ (error less than 0.5 \%). This inequality together with the
constraint $L < L_*$ implies $|y|^2=r<\frac{\epsilon^2 L_* \sin^2
\theta}{4 \gamma^2}$, so the supersymmetry-breaking parameter
$F=|Y|$ is automatically less than $\epsilon \sqrt{L_*}/2 \gamma$.
If we take the susy breaking cutoff $F_*$ greater than this
number, the integral \eq{nonsusyint} becomes independent of
$F_*$ and is given by
\begin{eqnarray}
 \CN_{2,con,M>0} &=& \int d^2\tau \, d^2
 v  \, \det g \, \nu(v), \\
 \nu(v) &=& \frac{16 \, \pi}{2 |\epsilon|^2} \int_0^\pi d\theta
 \int_0^{\frac{\epsilon^2 L_* \sin^2 \theta}{4 \gamma^2}} dr
 \int_{\frac{4 \gamma^2 r}{\epsilon^2 \sin^2 \theta}}^{L_*} ds \, r s
 \label{eq:nonsusyconifint}
 \\
 &=& \frac{3 \pi^2 L_*^4 \epsilon^2}{128 \, \gamma^4}.
\end{eqnarray}
Up to logarithmic factors, this has the same dependence on $v$ as
the number of supersymmetric near-conifold vacua with $M>0$, which
according to \eq{susyposMrho} is given by
\begin{equation}
 \nu_{susy}(v) = \frac{7 \, \pi^2 L_*^4 \epsilon^2}{8}.
\end{equation}
Nonsusy near-conifold vacua with $M>0$ are sparser than their susy
counterparts; their density ratio is $1/14 \gamma^4 \ll 1$.

\subsubsection{Distribution of cosmological constants in conifold
limit}

The distribution of cosmological constants for $M>0$ vacua near
the conifold point is obtained by adding $\delta(V-\Lambda)$ to
the integrand of \eq{nonsusyconifint}, with $V$ as given by
\eq{Vconif2}:
\begin{eqnarray}
 \nu(v,\Lambda) &=& \frac{16 \, \pi}{2 |\epsilon|^2} \int_0^\pi d\theta
 \int_0^{\frac{\epsilon^2 L_* \sin^2 \theta}{4 \gamma^2}} dr
 \int_{\frac{4 \gamma^2 r}{\epsilon^2 \sin^2 \theta}}^{L_*} ds \, r
 s \, \delta(r - \frac{3 \epsilon^2}{4} s - \Lambda)\\
 &=& \frac{16 \pi}{2 \epsilon^2} \left( \frac{4}{3 \epsilon^2}
 \right)^2 \int_0^\pi d\theta \int_0^M dr \, r(r-\Lambda)
\end{eqnarray}
where $M=\mbox{min}(- \frac{\Lambda \sin^2 \theta}{3
\gamma^2},\frac{3 \epsilon^2 L_*}{4} + \Lambda,\frac{\epsilon^2
L_* \sin^2 \theta}{4 \gamma^2})$. The first two entries in
$\mbox{min}(\ldots)$ come from the integration boundaries of $s$.
Since $s>0$, $M$ has to positive, and we get the following
condition on $\Lambda$ to get a nonzero result:
\begin{equation} \label{eq:nonsusyCCint}
 -\frac{3 \epsilon^2 L_*}{4} < \Lambda < 0.
\end{equation}
In particular we see that the cosmological constant can only be
negative for these vacua, so there are no meta-stable de Sitter
vacua in the conifold region either. This is because the $M>0$
condition forces the positive term in $V$ to be smaller than the
negative one. Under these conditions on $\Lambda$, we have in fact
that $M=- \frac{\Lambda \sin^2 \theta}{3 \gamma^2}$, except in a
very small interval (width of order $1/\gamma^2$ relative to
\eq{nonsusyCCint}) near the lower bound on $\Lambda$. Outside
that interval, the integral is therefore straightforward to
compute and equal to
\begin{equation}
 \nu(v,\Lambda) = \frac{8 \pi^2}{27 \gamma^4}
 \left(\frac{\Lambda}{\epsilon^2}\right)^3.
\end{equation}
Inside the small boundary interval, the density quickly drops to
zero.

\section{Finding vacua with quantized flux}

So far, we have been discussing the problem of finding the volume of
the region in ``flux space'' which contains vacua, not the problem of
counting physical vacua with quantized flux.  This approximation was
necessary at a very early stage, when we solved the conditions
$DW(z)=0$ or $V'(z)=0$ for some of the fluxes in terms of the others,
as at a generic point $z$ in moduli space these conditions will have
no solution with integer fluxes.

Nevertheless, we can get results for the problem with quantized fluxes
using these techniques.  The approach will be to consider a region $R$
in moduli space, and characterize the corresponding region $S_L$ in the
``space of fluxes'' $\IR^K$ which contains vacua with $N\eta N / 2 \le L$.
Since the number of flux
vacua in this region of moduli space is the number of points
with integral coordinates, {\it i.e.} points in $S_L \cap \IZ^K$,
we need ways to estimate this number from facts about the geometry of $S_L$.

The region $S$ containing supersymmetric vacua
is not hard to describe \cite{AD}.  
At a given point $(z,\tau)$ in moduli space, the $2n+2=K/2$ conditions 
$0=N^\alpha \Re D_i\Pi^\alpha=N^\alpha \Im D_i\Pi^\alpha$ 
pick out a subspace $S_{z,\tau}\cong \IR^{K/2}$ of the space of fluxes.
The region $S$ is then the union of the individual $S_{z,\tau}$
for all values of moduli $(z,\tau)\in R$.  One can show that the
subspaces vary transversally (the ``non-degeneracy condition'' of \cite{AD})
and thus this will fill out a $K$-dimensional region.
Finally, the subset $S_L\subset S$ is obtained by imposing the
additional constraint $N\eta N\le/2 L$.  

\EPSFIGURE{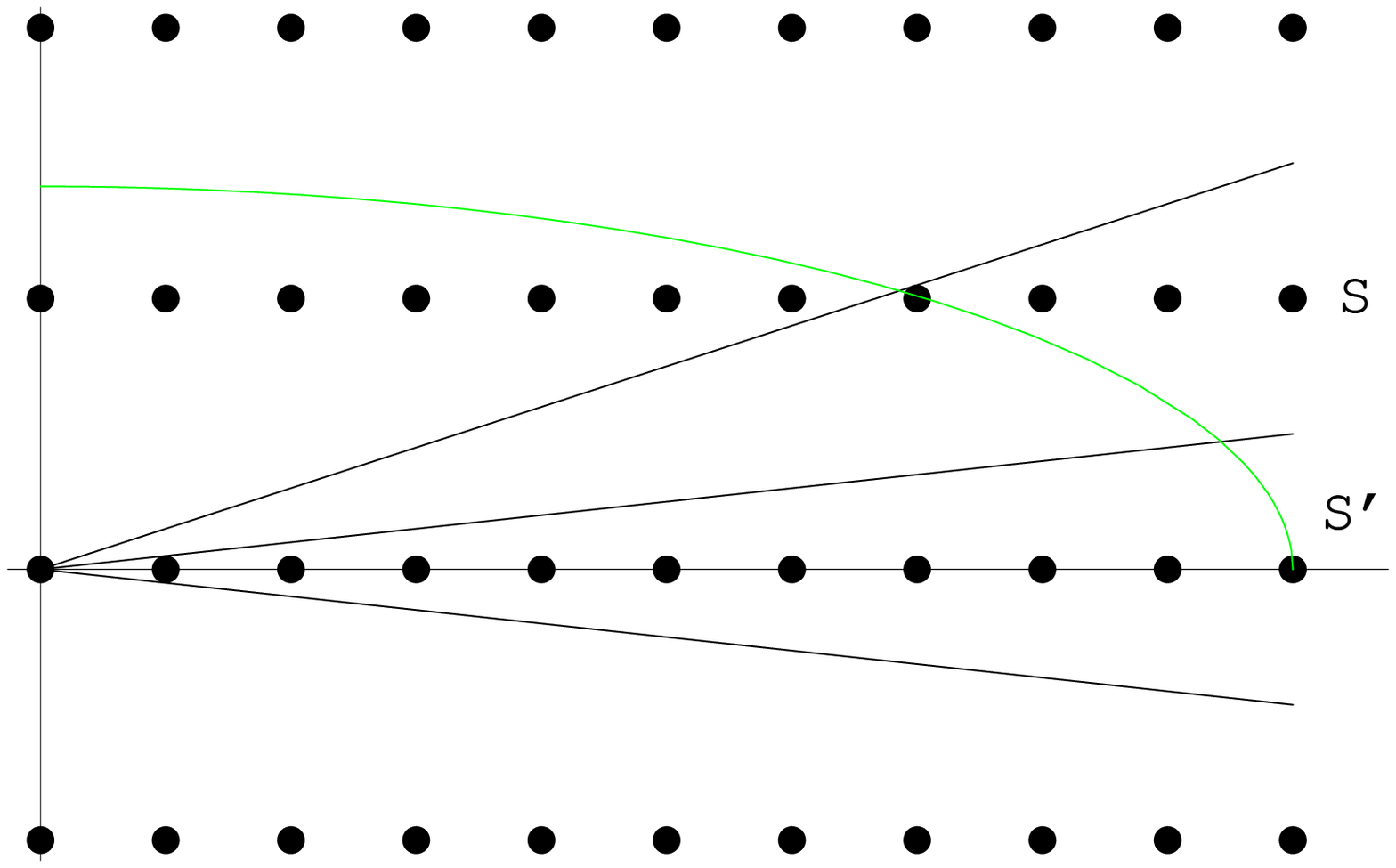,height=2.7in,angle=0,trim=0 0 0 0}
{Two possible regions $S$ and $S'$ containing supersymmetric vacua
in a $K=2$ flux space.  $S$ contains many fewer quantized flux points 
at small $L$ than its volume, while $S'$ contains many more.
\label{cones}}

It is clear that $S$ is a cone; in other words $\lambda S\cong S$.
Furthermore, for a small region $R$, the constraints $DW=0$ will not
vary much, so one can think of $S$ as roughly a cone over 
the product of a $K/2-1$-dimensional sphere 
(at fixed radius in $\IR^{K/2}$) with a $K/2$-dimensional ball.
Over a small region, the constraint $N\eta N/2\le L$ is not very different
from a positive definite quadratic constraint, so $S_L$ is roughly
the $r \le \sqrt{L}$ part of this cone.\footnote{
$S_L$ is however not convex, and
the theorems of Minkowski and
Mordell called upon in v1 of this paper are not applicable.}
For example, for $K=2$, one can have regions $S_L$ as pictured in figure 
\ref{cones}.

Having characterized $S_L$, our goal is to count how many lattice
points it contains.
The most basic estimate is of course that, if we make $L$ sufficiently
large, this number of lattice points will be well approximated by
the volume of $S_L$.  Furthermore, it is intuitively clear
(see \cite{GofN}, 2.iv for more precise statements) that the leading
correction to this is proportional to the surface area of the boundary
of $S_L$, so for large $L$
\begin{equation}\label{eq:Nasymp}
N(L) = L^{K/2} V(S_1) + L^{K/2-1/2} A(S_1) + {\cal O}(L^{K/2-1/2-\epsilon})
\end{equation}
where $V(S_1)$ and $A(S_1)$ are the volume and surface
area at $L=1$.

How big need $L$ be to reach this scaling regime?  In addition to
the condition $L^{K/2} V(S_1) >> 1$ for the region to contain many
lattice points, one clearly needs
\begin{equation}\label{eq:surface}
\sqrt{L} >> {A(S_1)\over V(S_1)}
\end{equation}
as well. 

To illustrate what happens for smaller values of $L$, consider the
case of $K=2$, and the two cones $S$ and $S'$ illustrated in figure
\ref{cones}.  Cone $S$, misaligned with the lattice, does not contain
any lattice points near the origin.
On the other hand, cone $S'$, aligned with the
lattice, contains roughly $r=\sqrt{L}$ (positive) lattice points
within a small distance $r$ from the origin, far more than its volume
$V \sim r^2\theta/2$ (where $\theta$ is the opening angle).

This phenomenon persists all the way out to $r \theta \sim 1$, at
which point the two dimensional nature of the cones starts to be
visible, and the estimate \eq{Nasymp} becomes valid.  This leads to
the condition $r > 1/\theta$.  Since $V(S_1)=\theta/2$
while $A(S_1) \sim 2$, this is the same as \eq{surface}.

Now the quantity $V(S_1)$ is just the integrated vacuum density over
the region $R$, and the quantity $A(S_1)$ can be computed as an
integral over moduli space using the same techniques.  In the
approximation that the vacuum density is just the volume form on
moduli space, the surface area of the boundary will just be the
surface area of the boundary in moduli space.  Taking the region
$R$ to be a sphere in moduli space of radius $r$, we find
$$
{A(S_1)\over V(S_1)} \sim {\sqrt{K}\over r}
$$
so the condition \eq{surface} becomes
\begin{equation}\label{eq:appcond}
L > {K\over r^2} .
\end{equation}

Thus, if we consider a large enough region, or the entire moduli space
in order to find the total number of vacua, the condition for the
asymptotic vacuum counting formulas we have discussed in this work to
hold is $L > c K$ with some order one coefficient.  But if we
subdivide the region into subregions which do not satisfy
\eq{appcond}, we will find that the number of vacua in each subregion
will show oscillations around this central scaling.  In fact, most
regions will contain a smaller number of vacua (like $S$ above), while
a few should have anomalously large numbers (like $S'$ above),
averaging out to \eq{Nasymp}.

\subsection{Flux vacua on rigid Calabi-Yau}

\EPSFIGURE{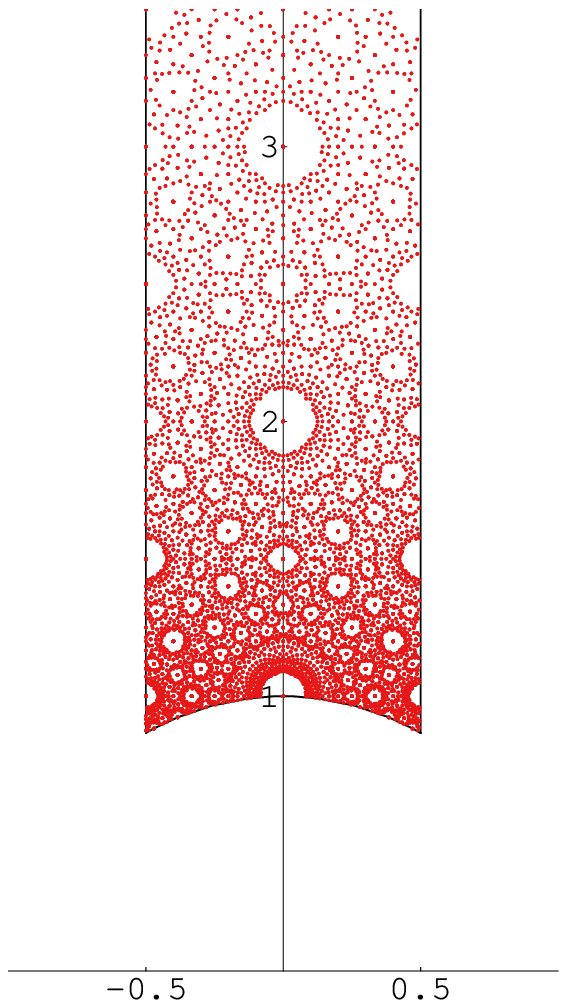,height=9.5cm,angle=0,trim=0 0 0 0}%
{Values of $\tau$ for rigid CY flux vacua with $L_{max} = 150$.\\
\\ 
 \label{T2vacua}}

As an illustration of this, consider the following toy problem with $K=4$,
studied in \cite{AD}.  The configuration space is simply the
fundamental region of the upper half plane, parameterized by $\tau$.
The flux superpotentials are taken to be
$$
W = A \tau + B
$$
with $A=a_1+ia_2$ and $B=b_1+ib_2$ 
each taking values in $\IZ+i\IZ$.  This would be obtained if
we considered flux vacua on a {\it rigid} Calabi-Yau, with no
complex structure moduli, $b_3=2$, and the periods $\Pi_1=1$ and $\Pi_2=i$.
The tadpole condition $N\eta N / 2 \leq L$ becomes
\begin{equation}\label{eq:simptad}
\Im A^* B \leq L
\end{equation}
One then has
\begin{equation}\label{eq:solvetau}
DW=0 \leftrightarrow \bar\tau = -{B\over A} .
\end{equation}
Thus, it is very easy to find all the vacua and the value of $\tau$ at
which they are stabilized in this problem.  We first enumerate all
choices of $A$ and $B$ satisfying the bound \eq{simptad}, taking one
representative of each orbit of the $SL(2,\IZ)$ duality group.  As
discussed in \cite{AD}, this can be done by taking $a_2=0$, $0\le b_1<a_1$ 
and $a_1 b_2 \leq L$.  Then, for each choice of flux, we take the
value of $\tau$ from \eq{solvetau} and map it into the fundamental
region by an $SL(2,\IZ)$ transformation. 
The resulting plot for $L=150$ is shown in figure \ref{T2vacua}.

A striking feature of the figure is the presence of holes around points such as $\tau = n i$ with $n \in \IZ$. 
At the center of such holes, there is moreover a big degeneracy of vacua. For example
there are 240 vacua at $\tau = 2i$. This clearly illustrates the phenomena discussed above. 
Starting with a tiny disk $D$ around $\tau = 2i$, the corresponding cone in flux space is very narrow, but
aligned with the lattice so it captures 240 points. For a somewhat displaced disk, the cone is not aligned
with the lattice and captures no points. Increasing the radius of the disk makes the cone wider, and at a certain
radius new lattice points enter. 

Despite the intricate structure of the finite $L$ result, it is true that a disc of
sufficiently large area $A$ will contain approximately $2 \pi A L^2$ vacua. 
This is illustrated for $L=150$ in figure \ref{discs}, where
estimated and real numbers of vacua are compared in discs around $\tau = 2i$ of stepwise increasing radius. 
Note that the first additional
vacua enter the circle at a coordinate radius $R \sim 0.12$, and that just beyond that radius, at $R \sim 0.15$,
the approximation becomes
good. The corresponding radius in the proper metric is $r \sim 0.04$. 
Since the holes around the integers clearly correspond to the worst case scenario for the estimate, we can 
thus conclude empirically that for $L = 150$, our estimate will be good when $r > 0.04$. 
Moreover, by comparing the results for different values of $L$, we found that the radii of the holes scale 
precisely as $1/\sqrt{L}$. Thus, our empirical result for the reliability of the approximation is 
$$
 r > {0.5 \over \sqrt{L}}, 
$$
which is compatible with \eq{appcond}.

\EPSFIGURE{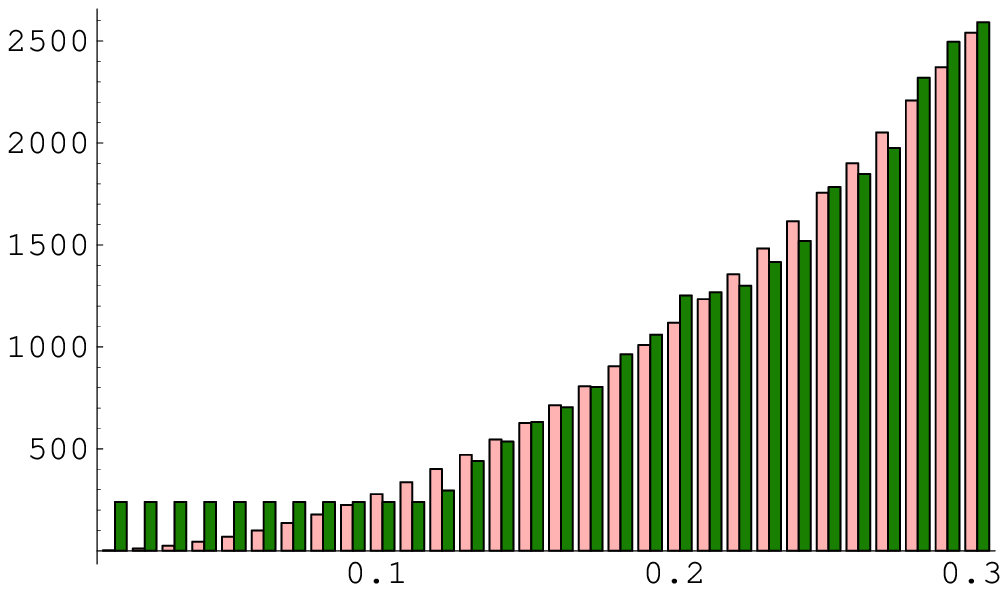,height=6cm,angle=0,trim=0 0 0 0}%
{Number of vacua in a circle of coordinate radius $R$ around $\tau = 2i$, with $R$ 
increasing in steps $dR = 0.01$. Pink bars give the estimated value, green bars 
the actual value. 
  \label{discs}}

\section{Conclusions}

We gave general results for the distribution of supersymmetric
and nonsupersymmetric flux vacua in type \IIb\ string theory, and
studied examples with one complex structure modulus in detail.
Let us conclude with a brief summary of the results, some comparisons
to what one might expect intuitively, and questions for further work.

A simplified picture of the results is that one can define an
``average density of vacua'' in the moduli space, which can be
integrated over a region of interest and then multiplied by a ``total
number of allowed choices of flux,'' to estimate the total number of
vacua which stabilize moduli in that region.  This estimate becomes
exact in the limit of large flux, and should be good for flux
satisfying the bounds discussed in section 5, say \eq{appcond} for
supersymmetric vacua.

The zeroth approximation for the average density
of vacua is the volume form on moduli space constructed from the
metric which would appear in the effective supergravity kinetic term;
explicitly if this is
$$
S_{moduli} = \int d^4x G_{ij}(z) \p z^i \p z^j ,
$$
then we find
$$
d\mu(z) \sim {\rm vol} = {(2\pi L)^{K/2}\over (K/2)!}
 {1\over (\pi M_p^2)^{n}} \sqrt{\det G(z)} d^{2n}z 
$$
in F theory and \IIb\ flux compactification, with $K\sim B_4$ fluxes
and tadpole bound $L$.

Of course, the actual results we obtained were of course more complicated,
depending not just on the metric but on curvatures and its derivatives.
However it is still useful to think of this ``zeroth approximation''
as the basic estimate, because in most of moduli space the
curvatures are proportional to the metric up to $O(1)$ coefficients.
In any case, this is the simplest distribution one can suggest which
uses no data other than the effective theory itself, and is thus the
``null hypothesis'' in this class of problem.

In comparing our actual results to this, perhaps the most striking
difference is the growth of vacuum density in regions of large
curvature, for example near conifold points.  The simple physical
explanation in this case is that we know that the structure of the
potential, or equivalently the dual gauge theory description of the
physics, can produce a hierarchically small scale $\Lambda$.  Since
the average spacing between vacua is expected to be $\Lambda$, we can
fit more vacua into such a region.

To make this quantitative, we need to understand the physics (or math)
which makes the number of vacua finite.  Here it is the tadpole
condition, schematically $AB=L$ where $A$, $B$ are two conjugate types
of flux, in dual terms controlling (say) the rank of the gauge group
and the gauge coupling.  A constraint $AB=L$ translates a distribution
$dA dB$ into a scale invariant distribution $L~dA/A$.

In the large complex structure limit, the superpotential goes
schematically as $A + B \tau^k$ for some power $k=1,2,3$.
This leads to vacua at $\tau^k\sim A/B$, and the distribution $dA/A$
translates into $d^2\tau/\Im\tau^2$, taking into account supersymmetry
which brings in the complexified modulus.  Such a measure is
naively scale invariant, and would lead to a diverging number of vacua.
But, it is important that the integration region for $\Re\tau$ is $[0,1)$,
so in fact the distribution falls off for large $\Im\tau$ and is integrable;
the total number of vacua with $|\tau|>T$ goes as $1/T$.
One can also understand this as a consequence of duality acting on the
other fluxes, which leaves finitely many inequivalent choices (as in the
toy example worked out in \cite{AD}).

Near a conifold, while the flux distribution is again roughly $L~dA/A$,
the dual gauge theory-type superpotential $W=A z + B z \log z$
leads to vacua at $\log z\sim A/B$ which have a
$|d\log\log z|^2 \sim d^2z/|z\log z|^2$ distribution.
In fact, the distribution
is rather similar to the previous one, with the identification $\tau=-\log z$.
Whether this has deep significance is not clear to us;
the large complex structure limit and conifold limit are not dual
and in other ways are not similar.  But it means that the number of vacua $N$
with a small scale $z < e^{-T}$, has a similar slow falloff, $N\sim 1/T$.

In some sense, many vacua are ``K\"ahler stabilized'' -- if the
K\"ahler potential were different, they would not exist.  An
illustration can be found in figure 4.  As usual, this is less
true for small $e^{\cal K}|W|^2$.

In section 5, we discussed the nature of the finite $L$ distribution
of vacua.  Because of flux quantization, this is far more complicated
than the smooth distributions which we have been computing.  On the
other hand, one can say a lot about it using the smooth distributions
and the methods we discussed.  The basic idea is to think of the
smooth distribution as computing a the volume of a region in flux
space which supports vacua.  Physical flux vacua satisfying flux
quantization are lattice points in this region.  While the volume of
the region controls the large $L$ asymptotics, its other characteristics
and in particular its surface area control the corrections to these
asymptotics.

In particular, the smooth distribution will always approximate the
finite $L$ distribution for sufficiently large $L$.  However, the
mimimal $L$ for this to be true depends on the size of the region in
moduli space in which one is counting vacua, and can be estimated as
$L \sim K/r^2$ for a ball of radius $r$.  Thus, the number of vacua in
a small region can deviate from the large $L$ distribution and even
the $L^{K/2}$ asymptotic for the total number of vacua, until $L$
becomes quite large.

Another way to say this is that at finite $L$, to get a smooth
density, one must average the number of vacua over regions in moduli
space of size $r \sim \sqrt{K/L}$.  The actual number of vacua in
smaller regions will show oscillations, with most regions having many
fewer vacua, and a few having many more vacua to make up for this.  We
discussed the simplest example in section 5, and a one complex modulus
example is discussed in \cite{GK2}.

It is not inconceivable that these considerations could significantly
decrease the total number of vacua in interesting examples with many
cycles, if the geometric factors such as total volumes of Calabi-Yau
moduli spaces were sufficiently small.  Work on computing these
volumes is in progress.

Let us move on to consider the results for specific types of vacua.
First, we give the ``null hypotheses'' or simplest pictures which one
might expect.  First, F-type nonsupersymmetric vacua should be
comparable in number to the supersymmetric vacua (given our
definitions, this is obvious for the D-type), up to factors like $2^n$
where $n$ is the number of moduli.  A heuristic argument for this is
that the potential $V(z)$ is quadratic in $W(z)$, so if $W$ had been a
polynomial of degree $d$, leading to $(d-1)^n$ vacua, then $V$ would
be a polynomial of degree $2d$, suggesting $(2d-1)^n$ vacua.  On
reflection, the main problem with this argument is that the equations
$V'=0$ are real equations which typically have fewer than $(2d-1)^n$
solutions.  Work on zeroes of randomly distributed polynomials with
natural geometric distributions \cite{Edelman} suggests that in some cases,
most zeroes are real, while in others there are many fewer real
zeroes, so this is inconclusive.  Finally, the metastability condition
might be expected to be weak, in the sense that if the masses of
bosons are distributed symmetrically about zero, the the fraction of
tachyon free vacua would be $2^{-2n}$.

In general, these expectations do hold, and the number of
non-supersymmetric vacua is comparable to the supersymmetric vacua.
However, the number of metastable nonsupersymmetric vacua falls off
drastically near conifold points, or more generally if the curvature
on moduli space becomes large.  This is rather surprising as naively
the curvature contribution to the mass matrix \eq{DcDV} goes the
other way; the conifold point has positive curvature which raises the
masses.  The explanation for the D type vacua is not complicated; it
has to do with the special form of the mass matrix which leads to a
``mixing'' with the dilaton-axion which forces a mode tachyonic, as
explained in 3.1.  Some experimenting with multimodulus models
suggests that this phenomenon is specific to one modulus and the
detailed form of this potential, again in the D breaking case.
We observed the same phenomenon for the F breaking; it may have
a similar explanation.

We found in section 3.3 that the distribution of values of  $e^{\cal
K}|W|^2$ or ``AdS cosmological constants'' is typically uniform
near zero, so that the fraction of vacua with $|\Lambda| <
\Lambda_{*}$ behaves as $\Lambda_{*}/L T_3$. A simple argument
for this was mentioned in \cite{durham}: for a given value of
moduli, the magnitude of $W$ is a single direction in ``flux
space,'' so the condition that it be small can be accomplished by
one tuning of fluxes.

However, the fact that this tuning can be made independent of the
choice of moduli is non-trivial.  In problems such as the attractor
mechanism with only one type of flux or charge for each cycle, it is
not true.  The heterotic string also has one type of flux per cycle
and is formally very similar; one cannot get small AdS cosmological
constant just from fluxes in this case.  Of course, there are many
more variables having to do with the gauge fields, which might make
small cosmological constant possible in this problem.

The simple argument also misses a good deal of structure in the
distribution; a good example is the behavior near the conifold point
displayed in figure 4.

The heuristic argument that ``nonsupersymmetric vacua are as common as
supersymmetric'' actually has a precise realization here, in the
existence of vacua with $W=D^2W=0$.  Since $W=0$ these are probably of
limited physical interest, in particular it is hard to see how to
stabilize their volume moduli.  Other nonsupersymmetric vacua exist
and are also roughly comparable in number.  Strangely enough, the
conditions of metastability and positive effective cosmological
constant were incompatible, at least in the conifold and large
structure limits.  If this were generally true, and true in
multi-modulus models, it would appear quite important.

Of course, all this was only a first cut at an actual counting of
vacua.  At this point it seems quite possible that many of the
nonsupersymmetric vacua which are stable under variations of the
moduli, have more subtle instabilities or inconsistencies.  This would
obviously be important to understand.

The scenario in which the numbers we are computing would be most
meaningful would be one in which such effects drastically reduced the
number of vacua, but in a way which was more or less uncorrelated to
the distributions we studied.  If we grant this, certain physical
expectations would result.  First, one finds that a high scale of
supersymmetry breaking is favored (of course, this could be the scale
in a hidden sector).  This is already true in the D breaking vacua,
for which at best the scale $M_{susy}^4$ is uniformly distributed, and
in some cases ({\it e.g.} near conifold points) even disfavors small
values.  While we did not find F breaking vacua with positive
cosmological constant, this may be an artifact of the one modulus case
(or of the special limits we considered).  But a high susy breaking
scale would be even more favored for F breaking vacua in models with
many moduli, as one expects the many supersymmetry breaking parameters
$F_i=D_iW$ to be roughly independent and uniformly distributed.

The expectation that a high scale of supersymmetry breaking would
lead to problems in tuning away the cosmological constant,
favoring a low scale of breaking, is probably not true for these
vacua.  The point is that the contribution to the energy which
does this tuning, the $-3e^{\cal K}|W|^2$ term or ``AdS
cosmological constant,'' gets contributions from many sectors of
the theory, including sectors with unbroken supersymmetry, and has
no {\it a priori} correlation to the supersymmetry breaking scale.
Given a particular supersymmetry breaking scale, this contribution
is then set by the requirement that $\Lambda\sim 0$, but since in
general vacua exist with a uniform distribution of this parameter,
this condition in itself does not favor any particular scale.

Thus, these distributions seem to favor ``moduli dominated gravity
mediated supersymmetry breaking,'' in the terminology of \cite{soft}.
One has a general prediction of a large gravitino mass, and risks the
common problems of supergravity mediated breaking, such as
non-universal soft breaking terms.  Perhaps this is less of a problem
in the D breaking models.

Finally, the moduli space measure (the metric on the space of metrics)
heavily disfavors large complex structure and large volume in the many
modulus models, as we will discuss in more depth in \cite{DD}.

\vskip 0.2in
{\it Acknowledgements}

We thank B. Acharya, B. Florea, S. Kachru, R. Kallosh, G. Moore,
B. Shiffman, S. Shenker, E. Silverstein, S. Thomas, P.M.H. Wilson,
E. Witten, S.-T. Yau and S. Zelditch for discussions and
correspondence.  We particularly thank S. Kachru for communications
regarding the work \cite{GK2} in progress, which stimulated us
to revise section 5.

This research was supported in part by DOE grant DE-FG02-96ER40959.
M.R.D. is supported by the Gordon Moore Distinguished Scholar
program at Caltech.

\end{document}